\documentclass[preprint,12pt]{elsarticle}
\usepackage[top=2cm, bottom=2cm, left=2cm, right=2cm]{geometry}
\usepackage{amsmath, amssymb,amsthm,color}
\usepackage{fancyhdr,url}
\usepackage{graphicx}
\usepackage{pbox}
\usepackage{array}
\usepackage{placeins}
\usepackage{bm}
\usepackage[title]{appendix}
\usepackage{algorithm,algorithmic,subfig}
\usepackage[english]{babel}
\usepackage{titlesec}
\newtheorem{theorem}{Theorem}[section]

\newtheorem{remark}{Remark}[section]

\usepackage{titlesec}
\titlespacing\section{0pt}{12pt plus 4pt minus 2pt}{0pt plus 2pt minus 2pt}
\titlespacing\subsection{0pt}{12pt plus 4pt minus 2pt}{0pt plus 2pt minus 2pt}
\titlespacing\subsubsection{0pt}{12pt plus 4pt minus 2pt}{0pt plus 2pt minus 2pt}

\graphicspath{{./}{./figs/}}

  \def\clap#1{\hbox to 0pt{\hss#1\hss}}

\usepackage{bm}
\providecommand{\mat}[1]{\bm{#1}}%
\renewcommand{\vec}[1]{\mathbf{#1}}
\newcommand{\vecalt}[1]{\bm{#1}}

\providecommand{\mA}{\ensuremath{\mat{A}}}
\providecommand{\mB}{\ensuremath{\mat{B}}}
\providecommand{\mC}{\ensuremath{\mat{C}}}

\providecommand{\mI}{\ensuremath{\mat{I}}}

\providecommand{\mM}{\ensuremath{\mat{M}}}

\providecommand{\mP}{\ensuremath{\mat{P}}}
\providecommand{\mQ}{\ensuremath{\mat{Q}}}
\providecommand{\mR}{\ensuremath{\mat{R}}}

\providecommand{\mU}{\ensuremath{\mat{U}}}
\providecommand{\mV}{\ensuremath{\mat{V}}}
\providecommand{\mW}{\ensuremath{\mat{W}}}

\providecommand{\mY}{\ensuremath{\mat{Y}}}

\providecommand{\mSigma}{\ensuremath{\mat{\Sigma}}}

\providecommand{\mPsi}{\ensuremath{\mat{\Psi}}}
\providecommand{\mPhi}{\ensuremath{\mat{\Phi}}}

\providecommand{\mPi}{\ensuremath{\mat{\Pi}}}

\providecommand{\vc}{\ensuremath{\vec{c}}}

\providecommand{\ve}{\ensuremath{\vec{e}}}

\providecommand{\vk}{\ensuremath{\vec{k}}}

\providecommand{\vn}{\ensuremath{\vec{n}}}

\providecommand{\vu}{\ensuremath{\vec{u}}}
\providecommand{\vv}{\ensuremath{\vec{v}}}

\providecommand{\vpi}{\ensuremath{\vecalt{\pi}}}
\providecommand{\vxi}{\ensuremath{\vecalt{\xi}}}
\providecommand{\vXi}{\ensuremath{\vecalt{\Xi}}}

\providecommand{\veps}{\ensuremath{\vecalt{\epsilon}}}

\newcommand{\sS}{\mathcal{S}}

\newcommand{\Exp}[1]{\mathbb{E}\left[#1\right]}
\newcommand{\parens}[1]{\left(#1\right)}
\newcommand{\braces}[1]{\left\lbrace #1 \right\rbrace}
\newcommand{\brackets}[1]{\left[ #1 \right]}

\newcommand{\Prob}[1]{\mathbb{P}\left[#1\right]}

\journal{Elsevier}

\begin{document} 
\begin{frontmatter}



\title{Sparse Polynomial Chaos Expansions via Compressed Sensing and D-optimal Design}


\author[]{Paul Diaz}
\ead{Paul.Diaz@colorado.edu}

\author[]{Alireza Doostan\corref{cor1}}
\ead{Alireza.Doostan@colorado.edu}

\author[]{Jerrad Hampton}
\ead{Jerrad.Hampton@colorado.edu}

\cortext[cor1]{Corresponding Author: Alireza Doostan}

\address{Ann and H.J. Smead Aerospace Engineering Sciences Department, University of Colorado, Boulder, CO 80309, USA}

\begin{abstract}
In the field of uncertainty quantification, sparse polynomial chaos (PC) expansions are commonly used by researchers for a variety of purposes, such as surrogate modeling. Ideas from compressed sensing may be employed to exploit this sparsity in order to reduce computational costs. A class of greedy compressed sensing algorithms use least squares minimization to approximate PC coefficients. This least squares problem lends itself to the theory of optimal design of experiments (ODE). Our work focuses on selecting an experimental design that improves the accuracy of sparse PC approximations for a fixed computational budget. We propose DSP, a novel sequential design, greedy algorithm for sparse PC approximation. The algorithm sequentially augments an experimental design according to a set of the basis polynomials deemed important by the magnitude of their coefficients, at each iteration. Our algorithm incorporates topics from ODE to estimate the PC coefficients. A variety of numerical simulations are performed on three physical models and manufactured sparse PC expansions to provide a comparative study between our proposed algorithm and other non-adaptive methods. Further, we examine the importance of sampling by comparing different strategies in terms of their ability to generate a candidate pool from which an optimal experimental design is chosen. It is demonstrated that the most accurate PC coefficient approximations, with the least variability, are produced with our design-adaptive greedy algorithm and the use of a studied importance sampling strategy.  We provide theoretical and numerical results which show that using an optimal sampling strategy for the candidate pool is key, both in terms of accuracy in the approximation, but also in terms of constructing an optimal design.
\end{abstract}
\begin{keyword}
Polynomial Chaos \sep Compressed Sensing \sep Optimal Design of Experiments \sep Subspace Pursuit; Coherence-optimal Sampling
\end{keyword}
\end{frontmatter}

%
%
\section{Introduction}\label{sec:intro}

Our understanding of complex scientific and engineering problems often stems from a general Quantity of Interest (QoI).  Practical analysis, design, and optimization of complex engineering systems  requires modeling physical processes and accounting for how uncertainties impact QoIs. Uncertainties may arise from variations in model inputs, measurements and data, or boundary and operating conditions. Much research has been done to quantify how the presence of uncertainty within a model manifests changes in a QoI \cite{ghanem2003stochastic,le2010spectral,xiu2010numerical}. This problem is often studied in the field of Uncertainty Quantification (UQ). 

A common approach in UQ for problems with random inputs involves expanding the QoI in a polynomial basis,  referred to as a \emph{polynomial chaos expansion} \cite{ghanem2003stochastic,xiu2002wiener}. One way to construct a PC expansion is to form a regression problem using Monte Carlo samples of the QoI. Often QoIs in scientific and engineering applications admit \textit{sparse PC expansions}, i.e., the QoI can be approximated by a small subset of the polynomial basis functions which capture important features of the model. This work focuses on QoIs which admit sparse PC expansions as detailed below. Sparsity may be exploited to regularize the regression problem; a concept studied in the context of compressed sensing \cite{candes2008introduction, donoho2006compressed,candes2006robust, elad2010sparse, eldar2012compressed}. In UQ, sparse PC expansions have been applied for a variety of different purposes \cite{doostan2011non,blatman2011adaptive,mathelin2012compressed,jones2015postmaneuver,sargsyan2014dimensionality,yan2012stochastic,yang2013reweighted,peng2014weighted,schiavazzi2014sparse,west2014uncertainty,jakeman2015enhancing,hampton2015coherence,bouchot2015compressed,peng2016polynomial,chkifa2016polynomial,winokur2016sparse,yang2016enhancing,adcock2017infinite,jakeman2017generalized}. 

Assume that the input parameters of our model are represented by a $d$-dimensional random vector $\vXi := (\Xi_1,\cdots,\Xi_d)$ with independent, identically distributed entries and with some joint probability density function $f(\vxi)$. We wish to approximate an unknown scalar QoI, with finite variance, denoted by $u(\vXi)$. Let $\psi_k(\vXi)$ represent a multivariate orthogonal polynomial, then we may write our QoI using a PC expansion as 
\begin{equation}\label{eq:pce}
u(\vXi) = \sum_{k=0}^\infty c_k \psi_k(\vXi).
\end{equation}
We truncate the expansion in \eqref{eq:pce} for computation, i.e., let $\vc = (c_1,\ldots,c_P)^T$ so that
\begin{equation}\label{eq:pce_truncation}
u(\vXi) = \sum_{k=1}^P c_k\psi_k(\vXi) + \epsilon(\vXi) \approx \sum_{k=1}^P c_k\psi_k(\vXi),
\end{equation}
where $\epsilon(\vXi)$ represents the \emph{truncation error} introduced by truncating the expansion to a finite number of terms. Often, in practice, many of the coefficients $c_k$ are negligible and thus $u(\vXi)$ admits a sparse representation of the form
\begin{equation}
u(\vXi) \approx \sum_{k \in \mathcal{C}} c_{k}\psi_{k}(\vXi),
\end{equation}
where the index set $\mathcal{C}$ has few elements, say $s=\vert \mathcal{C}\vert\ll P$, and we say that our QoI is \emph{approximately sparse} in the polynomial basis. 

The polynomials $\psi_k(\vXi)$ are selected with respect to the probability measure $f(\vxi)$ so that they are orthogonal, e.g., when $\vXi$ obeys a jointly uniform or Gaussian distribution (with independent components), $\psi_k(\vXi)$ are multivariate Legendre or Hermite polynomials, respectively \cite{xiu2002wiener}. We assume $\psi_k(\vXi)$ is obtained by the tensorization of univariate polynomials orthogonal with respect to the probability density function of the coordinates of $\bm\Xi$, and that $\psi_k(\vXi)$ is of \textit{total order} less than or equal to $p$. This formulation implies that there are $P := {{p+d}\choose{d}}$ basis polynomials. Furthermore, we assume that $\psi_k(\vXi)$ are normalized such that $\Exp{\psi_k^2}=1$, where $\mathbb{E}[\cdot]$ denotes the mathematical expectation operator. 

For $i=1,\ldots,N$, where $N$ is the number of independent samples considered, the computational model is evaluated for each realization of $\vXi$, which we denote $\vxi_i$, and yields a corresponding value of the QoI $ u(\vxi_i)$.  The coefficients $\vc$ are approximated using an experimental design consisting of samples $\braces{\vxi_i}_{i=1}^N$ and their corresponding QoIs $\braces{u(\vxi_i)}_{i=1}^N$, which are related by the linear system $\vu \approx \mPsi \vc$, where
\begin{equation}\label{samples}
\mPsi(i,j):= \psi_j(\vxi_i) \text{  and  } \vu:= \brackets{u(\vxi_1),\cdots,u(\vxi_N)}^T.
\end{equation}
Further, let $\mW$ be a diagonal positive-definite weight matrix such that $\mW(i,i)$ is a function of $\vxi_i$, which depends on the sampling strategy described in Section \ref{sec:sampling}.  Let $\mPhi := \mW\mPsi$ and $\vv := \mW\vu$. Under this sampling strategy, we consider the linear system 
\begin{equation}\label{eq:weighted_pce}
 \vv \approx  \mPhi \vc.
\end{equation}
In compressed sensing a sparse approximation $\hat{\vc}$ of $\vc$ is obtained by solving the optimization problem
\begin{equation}\label{eq:surrogate_approximation}
\hat{\vc} = \underset{\vc}{\text{argmin}} \| \vc \|_0 \quad \text{ subject to } \|\vv - \mPhi\vc\|_2  \leq \delta,
\end{equation}
where $\| \vc \|_0 = \braces{\text{the number of indices } k \text{ such that } c_k \neq 0}$ measures the sparsity of $\vc$. In (\ref{eq:surrogate_approximation}), $\delta$ is a tolerance of solution inaccuracy due to the truncation of the expansion. While, the problem (\ref{eq:surrogate_approximation}) is NP-hard to solve, approximate solutions may be obtained in polynomial time using a variety of greedy algorithms including \emph{orthogonal matching pursuit (OMP)} \cite{tropp2005signal,tropp2007signal,needell2010signal, davenport2010analysis}, \emph{compressive sampling matching pursuit (CoSaMP)} \cite{needell2009cosamp, pal2016stochastic}, and \emph{subspace pursuit (SP)} \cite{dai2009subspace}, or convex relaxation via $\ell_1$-minimization \cite{candes2008introduction, donoho2006compressed}. The key advantage of an approximation via compressed sensing is that, if the QoI is approximately sparse, stable and convergent approximations of $\vc$ can be obtained using $N < P$ random samples of $u(\vXi)$, as long as $\mPhi$ satisfies certain conditions~\cite{candes2008introduction,donoho2006compressed,doostan2011non,rauhut2012sparse,hampton2015coherence,adcock2017infinite}. Reducing the number $N$ of QoI samples, while maintaining solution accuracy and stability, is one of the main goals in UQ, and motivates the present study.

%
%

We use the linear system $\vv \approx \mPhi\vc$ to approximate the vector of coefficients $\vc$, by first generating $N$ samples of $\vXi$, $\braces{\vxi_i}_{i=1}^N$ namely, and simulating the corresponding QoIs $\braces{u(\vxi_i)}_{i=1}^N$.  Imagine that each simulation is costly, thereby limiting the number of simulations we are allowed to perform.  Under these circumstances, a natural question to ask is, can we choose our samples in a strategic way that improves the approximation of $\vc$? This paper aims to answer this question by focusing on the construction of the weighted measurement matrix $\mPhi$ for the purpose of solving the optimization problem in \eqref{eq:surrogate_approximation}. While our results are based on SP, the general ideas may be extended to, e.g., OMP or CoSaMP.

Let us briefly review the original SP algorithm (see Algorithm \ref{alg:SP}) with the specific focus of solving \eqref{eq:surrogate_approximation}. Define $\vv_p := \text{proj}(\vv,\mPhi_\sS) = \mPhi_\sS\mPhi_\sS^\dagger \vv$ to be the projection of $\vv$ onto the column space of $\mPhi_\sS$, where $\sS\subset \braces{1,\ldots,P}$ is an estimate of the support set of $\vc$, and $\mPhi_{\sS} := \mPhi(\;:\;,\sS)$, i.e., $\mPhi_\sS$ is the submatrix made from columns of $\mPhi$ corresponding to the set $\sS$. Further, define $\vv_r := \text{resid}(\vv,\mPhi_
\sS) = \vv - \vv_p$ to be the residual vector and $K \geq s $ to be an approximate upper bound on the number of non-zero coefficients in $\vc$. Note that in practice the optimal value of $K$ (or the tolerance parameter $\delta$) is not known {\it a priori} and should be estimated, for instance, using a cross-validation procedure \cite[Section~3.5]{doostan2011non} as discussed in Section \ref{subsec:cross-validation}. 

\begin{algorithm}[!h]                      
\caption{Subspace Pursuit (SP)}          
\label{alg:SP}                           
\textbf{Input:}  $K, \mPhi, \vv$\\
\textbf{Initialization:} \\
1) $\sS^0 = \lbrace K$ indices corresp. to the largest in $\vert\cdot \vert$ entries of the vector $\mPhi^T\vv \rbrace$. \\
2) $\vv_r^0 = \text{resid}(\vv,\mPhi_{\sS^0}).$\\  

\textbf{Iteration: }At the $\ell$th iteration perform the following steps:\\
1) $\tilde{\sS}^{\ell} = \sS^{\ell -1}\cup \lbrace K$ indices corresp. to the largest in $\vert \cdot \vert$ entries of the vector $\mPhi^T \vv_r^{\ell-1} \rbrace$.\\
2) Set $\hat{\vc} = \mPhi^\dagger_{\tilde{\sS}^{\ell}}\vv$.\\
3) $\sS^\ell = \lbrace K$ indices corresp. to the largest in $|\cdot |$ elements of $\hat{\vc} \rbrace$.\\
4) $\vv_r^\ell = \text{resid}(\vv,\mPhi_{\sS^\ell}). $\\
5) If $\ell = P$ quit iterating.\\
6) If $ \|\vv_r^\ell \| > \|\vv_r^{\ell -1}\|$ set $\sS^\ell = \sS^{\ell-1}$ and quit iterating. \\ 

\textbf{Output:} \\
1) The approximate PC coefficients $\hat{\vc}$, satisfying $\hat{\vc}(\braces{1,\ldots,P} \setminus \sS^\ell) = \bm{0}$ and $\hat{\vc}_{\sS^\ell} = \mPhi^\dagger_{\sS^\ell}\vv$.
\end{algorithm}

 A key step in SP algorithm (as well as in OMP and CoSaMP) involves solving an over-determined least squares problem corresponding to a subset of columns of $\mPhi$. In SP, this problem is represented by step 2) of the iteration. This over-determined least squares problem lends itself to optimization techniques from \emph{optimal design of experiments (ODE)} \cite{pukelsheim2006optimal,sinha2014optimal}. We propose the use of an \emph{alphabetic optimality criterion} from ODE to sequentially augment the experimental design according to the support set estimate $\sS$ on any given iteration, which determines the construction of the weighted measurement matrix $\mPhi$. The sequential augmentation is done such that once a sample is selected to be a member of the design it is never removed, i.e., samples and their corresponding QoI evaluations are never discarded while augmenting the experimental design. This constraint is necessary due to the computational cost of evaluating the QoI. 
 
Compared to PC approximations via over-determined {\it least squares approximation} (LSA), where the use of ODE has been well explored, ODE for compressed sensing has received less attention. The idea of using ODE for the purpose of compressed sensing, specifically when prior information about the sparsity is known in advance, is not necessarily new \cite{davenport2015constrained, chepuri2014compression}. For instance, in \cite{davenport2015constrained}, the authors show that improvements in signal reconstruction accuracy are possible under the assumption that the support (the locations of the non-zero entries) of a signal, with respect to a given basis, is either known or can be estimated. This was the case in their example dealing with Fourier measurements of Wavelet sparse signals \cite[Section~4]{davenport2015constrained}, where it was shown that using an alphabetic optimality criterion from ODE to sequentially choose measurements reduced the mean-squared error of signal reconstructions. However, the assumption that prior information regarding the support of  $\vc$ is known in advance is often too restrictive in the context of PC expansions. In fact, for most scientific and engineering applications none of the coefficients are zero and the locations of the largest in magnitude coefficients are unknown. Hence, the motivation for our approach. 
\subsection{Contributions of this paper}
We propose a {\it quasi-random} sampling strategy for building experimental designs which uses a studied optimal importance sampling and an alphabetic optimality criterion from ODE as described in Sections \ref{subsec:cos} and \ref{sec:doe}, respectively. This strategy is exploited by a modified SP algorithm which builds an experimental design sequentially and chooses input samples from a candidate pool while simultaneously constructing an approximation of the PC coefficients at each iteration. A similar idea was presented in \cite{fajraoui2017optimal}, where alphabetic optimal criteria from ODE were employed to sequentially build experimental designs with \emph{least angle regression (LAR)} for PC expansions. 

Our work differs from \cite{fajraoui2017optimal} where the authors concluded that sequential design augmentation using \emph{$D$-optimality} (an alphabetic optimal criterion) showed poor performance relative to other sampling techniques, whereas our results indicate otherwise. Further, our work is based on the SP algorithm rather than LAR. This work also introduces a novel method for sequential design augmentation based on the ideas presented in \cite{seshadri2016effectively}. The method uses a QR with column pivoting algorithm to construct and augment designs which allows for simple numerical implementation compared to existing greedy or exchange algorithms. We show that our modified SP algorithm reduces relative error and variability in the approximated PC coefficients compared to the standard SP algorithm. We provide evidence to support this claim in Section \ref{sec:numerics} by investigating manufactured sparse PC expansions with additive noise, a mathematical model for a Duffing oscillator, the Ishigami function, and a wing weight function. Finally, we provide theoretical and numerical results in Sections \ref{subsec:doe_theory} and \ref{sec:numerics}, respectively, which show that using an optimal sampling strategy for the candidate pool is beneficial, in terms of constructing an optimal design and the solution accuracy.

\section{Sampling}\label{sec:sampling}

In this section we outline the sampling method used in this work. First, we highlight some preliminary sampling definitions. Second, we consider sampling according to the random variables defined by the orthogonality measure $f(\vxi)$, which is often used in PC approximation and referred to as \textit{standard Monte Carlo sampling}; see \cite{hosder2006non,le2010spectral, doostan2011non, mathelin2012compressed}. Third, we outline the \emph{coherence-optimal} sampling strategy, as it has been shown to improve the stability and accuracy of over-determined least squares PC approximations relative to standard Monte Carlo sampling in either the Legendre or Hermite bases \cite{hampton2015coherence}. This discussion on sampling methods is motivated by the over-determined least squares problems solved within SP as mention in Section \ref{sec:intro} and specified in Step 2) of the iteration in Algorithm \ref{alg:SP}.

\subsection{Sampling Preliminaries}\label{subsec:preliminaries}

It is necessary here to outline some preliminary definitions for the topics in Section \ref{subsec:cos}. We follow the discussion and notation of \cite[Section~2.1]{hampton2015coherence} but limit our presentation to polynomials bounded over their domain of orthogonality measure. Extensions to unbounded polynomials, e.g., for Gaussian inputs and Hermite polynomials, can be found in \cite{hampton2015coherence}. Recall the set of basis polynomials $\{ \psi_k(\vxi)\}_{k=1}^P$ defined in Section \ref{sec:intro} and define $B(\vxi)$ to be 
\begin{equation}\label{eq:L2_coherence}
B(\vxi) := \sqrt{\sum_{k=1}^P |\psi_k (\vxi)|^2}. 
\end{equation}
Here, $B^2(\vxi)$ represents a uniformly least upper bound on the sum of squares of the basis polynomials considered. A bound on $B(\vxi)$ may be attained from 
\begin{equation}\label{eq:L2_coherence_bound}
B^2(\vxi) \leq P \; \underset{k=1:P}{\text{sup}} \; \vert \psi_k(\vxi)\vert^2,
\end{equation}
where bounds on $\text{sup}_{k=1:P} \; \vert \psi_k(\vxi)\vert^2$ are known for certain types of orthogonal polynomials \cite{hampton2015compressive,rauhut2012sparse,szeg1939orthogonal,dominici2007asymptotic,askey1965mean,muckenhoupt1970asymptotic,nevai1994generalized}. Hence, $\psi_k(\vxi)/B(\vxi) \leq 1$ and
\begin{equation}\label{eq:c}
c = \left( \int f(\vxi)B^2(\vxi) \right)^{-1/2}
\end{equation} 
is such that 
\begin{equation}\label{eq:c2}
c^2\int f(\vxi)B^2(\vxi)d\vxi = 1,
\end{equation}
and 
\begin{equation}\label{eq:Y_distribution}
f_{\mathbf{Y}}(\vxi) := c^2f(\vxi)B^2(\vxi),
\end{equation}
defines a probability density for the variable $\mathbf{Y}$. 

This formulation is designed to identify distributions for $\mathbf{Y}$. However, under these conditions we can no longer guarantee that $\Exp{\psi_i(\mathbf{Y})\psi_j(\mathbf{Y})} = \delta_{i,j}$, in which case the polynomials are not necessarily orthogonal. If we let 
\begin{equation}\label{weighting}
w(\mathbf{Y}) := \frac{1}{cB(\mathbf{Y})},
\end{equation}
then the weight function $w(\mathbf{Y})$ ensures that $\{w(\mathbf{Y})\psi_i(\mathbf{Y})  \}_{i=1}^P$ are orthonormal random variables. This construction motivates how we define the diagonal positive-definite weight matrix $\mW$ from \eqref{eq:surrogate_approximation}, i.e.,
\begin{equation}\label{weights}
\mW(i,i) = w(\vxi_i),
\end{equation}
where $\vxi_i$ is the $i$th realization of $\mathbf{Y}$.  We denote all realized random vectors by $\vxi$ without regard for the corresponding sampling distribution employed and we note that the weight function $w$ depends on the sampling strategy.

While our work focuses on compressed sensing, where $N < P$, we first highlight some important, overlapping concepts from LSA.  In LSA of PC expansions, typically $N \gg P$, and the approximated PC coefficients are given by
\begin{equation}\label{eq:LSA}
\hat{\vc} =  \underset{\vc}{\text{argmin}} \| \mPhi\vc - \vv\|_2 
\end{equation}
In this case, $\vc$ is computed by solving the system of normal equations
\begin{equation}\label{eq:normal}
\mPhi^T\mPhi\hat{\vc} = \mPhi^T\vv.
\end{equation}
Even though \eqref{eq:surrogate_approximation} differs from the LSA in \eqref{eq:LSA}, greedy methods like SP for solving \eqref{eq:surrogate_approximation} involve an over-determined LSA to construct $\hat{\vc}$. This fact motivates our interest in optimizing the LSA in SP. 

\subsection{Standard sampling}\label{subsec:standard}

Standard Monte Carlo sampling involves constructing the samples $\braces{\vxi_i}_{i=1}^N$ according to $f(\vxi)$, the orthogonality measure for a given PC basis.  This sampling strategy implies equal weighting, i.e., $w(\vxi) = 1$ and $\mW = \mI$. In the case of $d$-dimensional Legendre polynomials, the standard method is to sample independently from the uniform distribution on $[-1,1]^d$. In the case of $d$-dimensional Hermite polynomials, the standard method is to sample each of the $d$ coordinates as an independent standard normal random variable. 

\subsection{Coherence optimal sampling}\label{subsec:cos}

The coherence parameter $\mu$, defined in \cite{hampton2015coherence} as 
\begin{equation}\label{eq:coherence}
\mu(\mY) := \underset{\vxi}{\text{sup}}\sum_{j=1}^P| w(\vxi)\psi_j(\vxi)|^2,
\end{equation}
plays a key role in the stability and convergence of least squares PC approximation \cite{cohen2013stability,hampton2015coherence,cohen2016optimal}. A smaller $\mu$ results in more stable and accurate LSAs \cite[Theorem~2.1]{hampton2015coherence}. This result motivated the design of a random sampling strategy referred to as \emph{coherence-optimal sampling} \cite{hampton2015coherence}. Coherence-optimal sampling seeks to find a sampling measure to minimize $\mu$. This strategy involves sampling $\vxi$ according to the distribution defined by \eqref{eq:Y_distribution} for a normalizing constant $c=1/P$. We mention that, $f(\vxi)$ is the measure for which the basis polynomials $\psi_k(\vxi)$ are naturally orthogonal. We define the weight function as
\begin{equation}\label{eq:weights}
w(\vxi) := \frac{1}{B(\vxi)}.
\end{equation}
 In the case where LSA is applied to all $P$ basis funcitons, and with the weight function as in \eqref{eq:weights}, sampling from \eqref{eq:Y_distribution} leads to the minimum possible $\mu = P$ compared to any other sampling measure \cite[Theorem~3.3]{hampton2015coherence}. Coherence-optimal sampling is performed with a Markov Chain Monte Carlo (MCMC) sampler to minimize the coherence parameter defined by \eqref{eq:coherence}; see \cite[Section~4.3.1]{hampton2015coherence}. To generate coherence-optimal samples, we use the \texttt{Matlab} code available at \url{www.github.com/CU-UQ}.

The MCMC sampler must be given a proposal distribution. When $p \leq d$, we use as proposal distributions, standard normal in the case of Hermite polynomials, and uniform over $[-1,1]^d$ in the case of Legendre polynomials, where samples are drawn independently. When $p>d$, samples are independently drawn from a uniform distribution on a $d$-dimensional ball of radius $\sqrt{2}\sqrt{2p+1}$ for Hermite polynomials as in \cite[Section~3.2]{hampton2015coherence}, and a $d$-dimensional Chebyshev distribution for Legendre polynomials. For more information on these proposal distributions see \cite{hampton2016compressive,dominici2007asymptotic}. When the cost of evaluating the QoI is expensive, the construction of the samples $\{ \vxi_i \}_{i=1}^N$ is not typically a computational bottleneck. Hence, the extra cost of the MCMC sampling is justifiable. 

Coherence-optimal sampling ensures stable computation of $\hat{\vc}$ via \eqref{eq:normal}, with a number of QoI computations that depends linearly (up to a logarithmic factor) on the number of $PC$ coefficients, i.e., $N  \sim  \mathcal{O}(P \; \text{log}\;P)$ \cite[Theorem~2.2]{hampton2015coherence}. Further, as this approach was applied to various numerical examples, it was empirically demonstrated that coherence-optimal sampling leads to either similar or considerably more accurate LSAs in comparison to sampling from $f(\vxi)$ \cite{hampton2015coherence, hadigol2017least}. In Section \ref{subsec:msf}, we also demonstrate this improvement in accuracy for the case of $N<P$ when using SP.

\section{Optimal design of experiments}\label{sec:doe}

Constructing a surrogate model requires sampling the input parameter space and performing experiments, whether physical or computational. The planning of this experimental procedure prior to conducting the experiment is referred to as \emph{design of experiments} (DoE). Often, experiments are expensive and time-consuming and inputs should be selected in order to extract as much information as possible for a given amount of experimental effort, this is the study of ODE. Historically, work on ODE dates back to 1918 \cite{smith1918standard}, which was expanded upon a few decades later in \cite{kiefer1985collected}. ODEs are commonly used in the context of least squares regression  \cite{box2005statistics,atkinson2007optimum,fedorov1972theory,fedorov2012model,pukelsheim2006optimal}. 
For a brief review and interpretation of a major class of ODE, known as \emph{alphabetic optimal design}, related to least squares PC approximation, the interested reader is referred to \cite[Section~4.5]{hadigol2017least}. 

In this work, we seek to construct surrogate models in the context of sparse PC expansions. This construction is performed using a greedy compressed sensing algorithm, and a key feature of this algorithm is the solution of over-determined least squares sub-problems. Our approach is to apply ODE to these sub-problems. To explain this ODE strategy, in Section \ref{subsec:D-optimal} we briefly review the $D$-optimality criterion, an alphabetic optimality criterion that is widely used in ODE and exclusively focused on in this work. Section \ref{subsec:dopt_construction} describes some conventional methods used to construct $D$-optimal designs. The primary method for constructing $D$-optimal designs in this work is a QR factorization with column pivoting algorithm which is outlined in Section \ref{subseq:RRQR}. We conclude by providing some theoretical results relevant to $D$-optimal designs in Section \ref{subsec:doe_theory}.

\subsection{D-optimal designs}\label{subsec:D-optimal}

Typically in ODE, the design points $\braces{\vxi_i}_{i=1}^N$ are chosen according to an alphabetic optimality criterion, which is a scalar function $\phi(\mM)$ of the so-called \emph{information matrix} $\mM$ defined as
\begin{equation}\label{eq:weighted_information_matrix}
\mM := \frac{1}{N}\mPhi^T\mPhi.
\end{equation}
The matrix $\mM$ plays an important role in the stability of LSAs, described by its deviation from the identity matrix \cite{hampton2015coherence, hadigol2017least}. An important observation is that $\mM$ does not depend on any realized values of the QoI $u$, and this means that different designs may be compared in terms of $\phi$ to judge their relative optimality prior to any simulations of the QoI.
\par
$D$-optimal (or determinant optimal) designs are obtained by maximizing the determinant of the information matrix, i.e., maximizing
\begin{equation}\label{eq:D-optimality}
\phi_D(\mM) :=  |\mM|^{1/P},
\end{equation}
where the $1/P$ factorization is a convenient normalization that has been used in the literature. $D$-optimal designs fall into the category of estimation-oriented optimal designs \cite{jones2012optimal}, which focus on the precise estimation of the coefficients $\vc$, thereby improving surrogate accuracy. An equivalent formulation involves minimizing the determinant of the inverse information matrix, i.e., minimizing $|\mM^{-1}|^{1/P}$  \cite{fedorov2012model}. 
\begin{remark}
 In \cite{fajraoui2017optimal},  $D$-optimal designs were employed in the context of sequential sparse PC approximation, and compared to a closely related criterion based on the objective function 
 \begin{equation}\label{eq:S-optimality}
 \phi_S(\mPhi) := \parens{\frac{\sqrt{\vert \mPhi^T \mPhi  \vert }}{\prod_{i=1}^P \|  \mPhi^{(i)}  \|_2}}^{1/P},  
 \end{equation} 
where $\mPhi^{(i)}$ represents the $i$th column of $\mPhi$.  The $S$-optimality criterion given by \eqref{eq:S-optimality}, was originally presented in \cite{shin2016nonadaptive} as a method of point selection for LSA. 
 \end{remark}

\subsection{Construction of alphabetic optimal designs}\label{subsec:dopt_construction}

Unlike the random sampling strategies of Sections \ref{subsec:standard} and \ref{subsec:cos}, ODE offers {\it deterministic} sampling methods to improve the PC approximation of $u$. In general, however, the alphabetic optimal designs are constructed by generating, either randomly or deterministically, a large number of candidate samples such that the selected optimal design depends on the choice of candidates. In this regard, we here consider ODE as {\it quasi-random} sampling. As we shall justify in Section \ref{subsec:doe_theory}, this work proposes using coherence-optimal sampling to generate candidate samples for ODE.

Let $M>N$ and $\braces{\vxi_i}_{i=1}^M$ be a pool of candidate samples generated with respect to the orthogonality measure $f(\vxi)$ in the case of standard Monte Carlo sampling, or \eqref{eq:Y_distribution} in the case of coherence-optimal sampling. Let $\mPsi_c$ represent the matrix of basis polynomials corresponding to the candidate samples $\braces{\vxi_i}_{i=1}^M$ and let $\mW_c$ be the appropriate weight matrix, then we define the candidate measurement matrix as $\mPhi_c := \mW_c\mPsi_c\in \mathbb{R}^{M \times P}$.  As previously mentioned, the information matrix does not involve evaluating the QoI $u$. This fact implies that when the computational bottleneck is the evaluation of the QoI for any given realization $\vxi$, the additional computational cost of constructing an optimal design is justifiable. Hence, this pre-processing of the candidate sample pool can improve PC approximation in a cost-effective manner. 

The problem of finding an exact $D$-optimal design, i.e., choosing $N$ out of $M$ rows of $\bm\Phi_c$ maximizing $\phi_D$, is NP-hard, and this fact has motivated the development of relaxation techniques. Two common methods for constructing alphabetic optimal designs are exchange \cite{mandal2015algorithmic, smucker2010design, fedorov1972theory, cook1980comparison, mitchell1974algorithm, wynn1970sequential, johnson1983some, atkinson1989construction, meyer1995coordinate} and greedy \cite{dykstra1971augmentation,song2009netquest,shin2016nonadaptive,gammerman2016conformal} algorithms. Heuristic exchange algorithms were among the earliest search methods proposed for the construction of optimal designs \cite{mandal2015algorithmic, smucker2010design}. These exchange algorithms were developed originally for $D$-optimal designs because they were computationally more feasible in comparison to other criteria \cite{fedorov2012model}, and it has been shown that $D$-optimal designs perform well compared to other criteria \cite{atkinson2007optimum}. For comparisons of performance between different exchange algorithms see \cite{cook1980comparison,johnson1983some,pronzato2008optimal,nguyen1992review}.  Greedy algorithms such as \cite[Algorithm~1]{hadigol2017least} involve starting with a random seed, i.e., a random row of the candidate matrix, then iteratively and exhaustively searching the entire remaining candidates to build the experimental design row-by-row. Both exchange and greedy algorithms involve exhaustive searches of the candidate matrix at each iteration and can be computationally expensive when $M$ is large. 
To avoid the computational cost of exhaustive searches, we instead employ another greedy approach based on QR factorizations with column pivoting to build $D$-optimal designs which is discussed further in Section \ref{subseq:RRQR}. 

\subsection{QR factorization with column pivoting on $\mPhi_c^T$}\label{subseq:RRQR}

In this work, we use QR factorizations with column pivoting to construct $D$-optimal designs. This idea is based on the work of \cite{sommariva2009computing} who originally used QR factorization with column pivoting to approximate Fekete points on compact multivariate domains, and more recently \cite{seshadri2016effectively}, where QR factorizations with column pivoting were used to sub-sample design points from a tensor grid for generating least squares polynomial approximations.  In this section, we briefly review QR factorization with column pivoting and the subsampling method proposed in \cite{seshadri2016effectively} with the specific purpose of subsampling rows of $\mPhi_c$ to generate $D$-optimal designs. 

QR factorization with column pivoting is a greedy heuristic commonly used for solving rank deficient problems. This method works by (i) determining the numerical rank $r < M$ of a $P \times M$ matrix, and (ii) permuting the columns of the matrix such that the first $r$ columns are linearly independent \cite{hansen2012least,golub2012matrix, gu1996efficient}. For this reason, QR with column pivoting is sometimes referred to as \textit{rank revealing QR} (RRQR). Consider applying this heuristic to sub-select rows of $\mPhi_c$. Let $\mPhi_c^T \in \mathbb{R}^{P \times M}$, then there exists a RRQR factorization 
\begin{equation}\label{eq:RRQR_psi_c}
\mPhi_c^T \mP  = \mQ(\mR_1 \; \; \mR_2),
\end{equation}
where $\mQ \in \mathbb{R}^{P \times P}$ is orthogonal; $\mR_1 \in \mathbb{R}^{P \times P}$ is nonsingular, well conditioned, and upper-triangular; and $\mR_2 \in \mathbb{R}^{P \times (M-P) }$ \cite{gu1996efficient}. In \eqref{eq:RRQR_psi_c}, $\mP \in \mathbb{R}^{M \times M}$ is a permutation matrix that permutes the columns of $\mPhi_c^T$ such that the absolute value of the diagonal entries of $\mR_1$ are in descending order. Let $\vpi$ be a vector that converts the pivots encoded in the matrix $\mP$ to the specific rows of $\mPhi_c$, i.e., $\vpi$ determines which rows to select from $\mPhi_c$ with
\begin{equation}\label{eq:pi}
\vpi := \mP^T\vn,
\end{equation} 
where $\vn := (1,2,\ldots,M)^T$ and the vector $\vpi$ contains ordered indices corresponding to rows of $\mPhi_c$ that are subselected via the RRQR factorization. Let $\vpi_N = \vpi(1:N)$ be the first $N$ entries of $\vpi$, then we define the $N$-point, $D$-optimal design as
\begin{equation}\label{eq:rrqr_psi}
\mPhi_N =  \mPhi_c(\vpi_N, \;:\;),
\end{equation}
where $\mPhi_N \in \mathbb{R}^{N \times P}$ is the submatrix of $\mPhi_c$ constructed by selecting the rows of $\mPhi_c$ indexed by $\vpi_N$.\

\textit{Subset selection} is a similar process where the aim is to produce a well-conditioned submatrix of $\mPhi_c$ with linearly independent rows. This process can produce a submatrix with smaller condition number compared to that given by RRQR alone \cite{seshadri2016effectively, golub2012matrix}. Subset selection can be accomplished in two steps. The first step is to compute the singular value decomposition (SVD) of $\mPhi_c^T = \mU\mSigma \mV^T$. The second step is to compute the RRQR decomposition of the transpose of the first $P$ right-singular vectors of $\mPhi_c$, i.e.,
\begin{equation}\label{eq:right_singular_vectors}
\mV(\;:\;,1: P)^T \mP^T = \mQ\mR,
\end{equation}
where the columns of $\mP^T$ encode the permutations. Once this computation is performed, equations \eqref{eq:pi} and \eqref{eq:rrqr_psi} may be used with $\mP = \mP^T$ to determine $\mPhi_N$. It should be mentioned, however, that subset selection is more expensive as it requires both an SVD and QR computation. The additional cost is justified when the QoI evaluations are computationally expensive. 

\subsection{Theory relevant for the design of experiments}\label{subsec:doe_theory}

We conclude this section by proposing two key heuristics related to the $D$-optimality criterion. The first heuristic we consider is that a candidate set generated with coherence-optimal samples, as described in Section \ref{subsec:cos}, will likely produce designs with larger values of $\phi_D$ compared to a candidate set that is constructed via standard Monte Carlo sampling. This is demonstrated empirically in Section \ref{subsec:doe_testing}. We justify the use of coherence-optimal samples relevant to the $D$-optimality of arbitrary sub-matrices of $\bm{\Phi}_c$ corresponding to $K$ basis functions. As we shall explain in Section \ref{subsec:DSP}, our modified SP algorithm selects designs from such sub-matrices of $\bm{\Phi}_c$. To investigate $D$-optimality of these matrices, and following \cite{hampton2017basis}, we consider a coherence parameter
\begin{equation*} 
\mu_K := \underset{\vxi}{\text{sup}} \sum_{j\in\mathcal{S}} | w(\vxi)\psi_j(\vxi)|^2,
\end{equation*}
associated with any set $\mathcal{S}\subset\{1,\dots,P\}$ of size $K=\vert \mathcal{S}\vert$. We mention that $\mu_K$ does not exceed the coherence parameter associated with the $P$ basis polynomials, i.e., $\mu_K \leq \mu$ where $\mu$ is defined as in \eqref{eq:coherence}. This bound on $\mu_K$ is a consequence of \cite[Theorem~3.3]{hampton2015coherence}.
\begin{theorem}\label{thm:stochastic_dominance}
Consider a basis of $K$ polynomials. Let $\mM \in \mathbb{R}^{K\times K}$ be an information matrix associated with a set of $M>K$ coherence-optimal samples, and their associated weights, with a coherence $\mu_K$. For $t \in (0,1),$
%
\begin{equation}\label{eq:phi_D_bound}
\Prob{\phi_D(\mM)\leq t } \leq 2K\exp\brackets{-cM\mu_K^{-1}(1-t)},
\end{equation}
where $\phi_D$ is defined in \eqref{eq:D-optimality}.
\end{theorem}

\noindent{\it Proof:} See Appendix \ref{appendix:proofs} for a proof of this theorem.

\begin{remark}
We emphasize that the coherence parameter $\mu$ does depend on the $P$ polynomial basis functions. Particularly, for the SP algorithm, the set of $K$ polynomials is not determined at the time of sampling, so the coherence in \eqref{eq:phi_D_bound} is not known {\it a priori}. However, the bound on the probability given by \eqref{eq:phi_D_bound} decreases exponentially as $\mu_K$ decreases. This decay is significant because the coherence-optimal sampling strategy of Section \ref{subsec:cos} specifically minimizes the coherence parameter $\mu$, which in turn bounds $\mu_k$. 
\end{remark}
The second heuristic is that the RRQR method described in Section \ref{subseq:RRQR} identifies a subset of samples to form a design such that the corresponding information matrix has a large determinant. Consider the partial QR factorization $\mPhi_c^T\mP = \mQ\mR$ and let $P$ be the rank of $\mPhi_c$, where
  \begin{equation}\label{eq:R}
 \mR = \parens{\begin{array}{c c}
 \mA & \mB \\
  & \mC\\
 \end{array} },
 \end{equation}
 denotes the triangular matrix in the partial QR factorization of $\mPhi_c^T$, such that $\mA$ has positive diagonal elements.  Let  $(\cdot)_{i,j}$ identify the $i$,$j$th entry of a matrix. The second heuristic is justified in part by \cite[Lemma~3.1]{gu1996efficient}, which we restate here as Theorem \ref{thm:rrqr}.
 \begin{theorem}\label{thm:rrqr}
 \cite[Lemma~3.1]{gu1996efficient} If $\bar{\mR}$ denotes the triangular matrix after a pivot exchange that interchanges the $i$th and $(j+P)$th columns of $\mR$ such that
 \begin{equation}\label{eq:p_k(R)}
 \bar{\mR} = \parens{\begin{array}{c c}
 \bar{\mA} & \bar{\mB} \\
 & \bar{\mC}
 \end{array}},
 \end{equation}
 then
 \begin{equation}\label{eq:det_ratio}
 \frac{\det(\bar{\mA})}{\det(\mA)} = \sqrt{(\mA^{-1}\mB)_{i,j}+ (\Vert\mC(:,j)\Vert_2\Vert\mA^{-1}(i,:)\Vert_2)^2 }.
 \end{equation}
 \end{theorem}
 Theorem \ref{thm:rrqr} shows that a RRQR algorithm, such as \cite[Algorithm~3]{gu1996efficient}, actively insures that the $\det(\bar{\mA})$ is monotonically increasing with repeated permutations, as the exchanges done for pivoting at each iteration may be chosen to maximize \eqref{eq:det_ratio}. Consequently, Theorem \ref{thm:rrqr} shows that the columns of $\mPhi_c^T$ may be selected by the RRQR pivots as in \eqref{eq:rrqr_psi} to produce a design with non-decreasing values of $\phi_D$. Notice that as RRQR is a greedy algorithm and only achieves a local minima. That is, upon completion of the column pivoting, Theorem \ref{thm:rrqr} does not necessarily guarantee that the constructed $D$-optimal designs have larger value of $\phi_D$  than a design constructed by randomly selecting $N$ columns of $\mPhi_c^T$. However, in Section \ref{sec:numerics}, we do show that the PC approximations corresponding to designs selected by RRQR via \eqref{eq:rrqr_psi} consistently outperform approximations where designs are constructed by randomly selecting columns of $\mPhi_c^T$, for a variety of QoIs. We also mention that a RRQR algorithm such as \cite[Algorithm~3]{gu1996efficient} may be used to efficiently construct $D$-optimal designs as it requires in the worst case scenario $\mathcal{O}(MP^2)$ floating-point operations.

%
 
%
%
\section{Improving Subspace Pursuit}\label{sec:subspace_pursuit}

In this section, we propose two methods for improving PC approximation with SP.  These methods are investigated numerically in Section \ref{sec:numerics}. The first method focuses exclusively on the sampling strategy employed to select input samples for the standard SP algorithm. The second method is the main contribution of this work and uses the same sampling strategy as the first but involves modifying the standard SP algorithm to sequentially construct an experimental design based on the current PC approximation given at any iteration of the algorithm, as opposed to constructing the entire design before executing SP. 

The motivation for building an experimental design sequentially is simple. The sparsity support set $\sS \subset \braces{ 1,\ldots,P }$ of $\vc$ is {\it a priori} unknown, but is needed for experimental design. In many iterative compressed sensing algorithms, an estimate of $\sS$ is obtained at each iteration, which can be used to generate, or more precisely augment, a design. Some examples of these algorithms include OMP, CoSaMP, and SP \cite{tropp2005signal, needell2009cosamp, dai2009subspace}. Ideally, $\sS$ would contain the indices corresponding to the largest, in magnitude, coefficients in $\vc$. In practice, once a sample point is added to the experimental design and its corresponding QoI has been computed, it should not be removed as it provides valuable information regarding the QoI. For computationally expensive models, we aim to show that it is advantageous to start with a relatively small experimental design to construct a PC approximation, then sequentially add new samples to the design and update the approximation. To perform this task we introduce a sequential strategy which augments the experimental design based on the current estimated support set $\sS$. 

This portion of the paper is organized as follows. In Section \ref{subsec:SP_optimal_sampling}, we introduce an improved sampling strategy for SP. In Section \ref{subsec:RRQR_adapt} we present a QR-based approach for sequential design augmentation. In Section \ref{subsec:DSP}, we outline the modified SP algorithm with sequential sampling. Finally, in Section \ref{subsec:cross-validation}, we describe a simple cross-validation procedure for $K$, an approximate upper bound on the sparsity of $\vc$, which is necessary for the numerical experiments presented in Section \ref{sec:numerics}.
\subsection{Subspace pursuit with $D$-coherence-optimal sampling}\label{subsec:SP_optimal_sampling}
The coherence-optimal sampling strategy presented in Section \ref{subsec:cos} is known to produce at least as accurate as PC approximations compared to standard Monte Carlo sampling from $f(\vxi)$ for both LSA and $\ell_1$-minimization \cite{hampton2015coherence,hampton2015compressive, hadigol2017least,alemazkoor2017near,hampton2017basis}. Further, constructing $D$-optimal designs from a large pool of candidate samples, regardless of how the candidate samples are generated, can improve least squares PC approximation accuracy compared to designs constructed randomly \cite{jones2012optimal, hadigol2017least}. In this work, we combine these two sampling strategies to create a random, then deterministic sampling strategy which we call \emph{D-coherence-optimal sampling}. Specifically, $D$-coherence-optimal sampling involves first generating a large pool of coherence-optimal samples by sampling from \eqref{eq:Y_distribution}, then constructing an $N$-point, $D$-optimal design from the candidate pool according to \eqref{eq:rrqr_psi}. In Section \ref{sec:numerics}, we demonstrate that SP with $D$-coherence-optimal sampling outperforms coherence-optimal sampling in a variety of problems.
\subsection{Design adaptation using RRQR}\label{subsec:RRQR_adapt} 
As mentioned in the introduction to this section, once a sample point is added to an experimental design and its corresponding QoI value has been computed, the sample point should not be removed from the design as it provides valuable information regarding the QoI. At each iteration of SP, an approximate coefficient vector $\hat{\vc}$ is obtained via LSA; see step 2) of the iteration in Algorithm \ref{alg:SP}. This LSA is computed by selecting columns of $\mPhi$ which are most correlated, in magnitude, with the residual vector; see step 1) of the iteration in Algorithm \ref{alg:SP}. The indices of these columns are represented by the estimated support set $\sS \subset \braces{1,\ldots,P}$, where $\vert \sS \vert = K$.  The goal of design adaptation is to add a sample point from the candidate pool to the experimental design such that the sample is chosen via $D$-optimality from the set $\braces{\vxi_i}_{i=1}^M$ according to $\sS$, which is equivalent to selecting a row from $\mPhi_c(\;:\;,\sS)$. It is important to point out that instead of considering the full candidate matrix, the adaptation considers the submatrix consisting of columns of $\mPhi_c$ corresponding to $\sS$. This process is performed such that each point in the final experimental design is a unique member of the candidate pool $\braces{\vxi_i}_{i=1}^M$. In this section, we outline a QR-based approach for performing the design adaptation we have described. Because this approach uses RRQR, it avoids calculating determinants explicitly which increase rapidly for large values of $N$ and $K$, and can be problematic for constructing $D$-optimal designs, as described in Section \ref{subsec:dopt_construction}. Instead of computing determinants explicitly, relative changes in determinants as in \eqref{eq:det_ratio}, which are less expensive to compute, are all that is required. 

Let $\mPhi_N  = \mPhi_c(\vpi_N,\;:\;)$ be an $N$-point, $D$-optimal design as in \eqref{eq:rrqr_psi} for some candidate matrix $\mPhi_c$, and let $\tilde{\mPhi}_N := \mPhi_N(\;:\;,\sS)$ be the submatrix of $\mPhi_N$ which is constructed from the columns of $\mPhi_N$ indexed by $\sS$. Similarly, let $\tilde{\mPhi}_c \in \mathbb{R}^{M \times K} := \mPhi_c(\;:\;,\sS)$ be the submatrix of $\mPhi_c$ which is constructed from the columns of $\mPhi_c$ indexed by $\sS$. For our design adaptation, it is necessary to construct an approximation of $\tilde{\mPhi}_c^T$ using $\tilde{\mPhi}_N^T$. We may write this approximation as 
\begin{equation}\label{eq:Psi_c_low-rank}
\tilde{\mPhi}_N^T \mPi \approx \tilde{\mPhi}_c^T,
\end{equation} 
where $\mPi \in \mathbb{R}^{ N \times M }$ and rank$(\tilde{\mPhi}_N^T \mPi) = $ min$(N,K)$. When using SP, $N \geq 2K$ \cite{dai2009subspace}, therefore rank$(\tilde{\mPhi}_N^T \mPi) = K$. The matrix $\mPi$ may be solved for using least squares regression where $\mPi = (\tilde{\mPhi}_N^T)^\dagger \tilde{\mPhi_c}^T$. Now, let the RRQR factorization of  $\tilde{\mPhi}_c^T-\tilde{\mPhi}_N^T\mPi$ be
\begin{equation}\label{eq:RRQR_adapt}
(\tilde{\mPhi}_c^T-\tilde{\mPhi}_N^T \mPi)\tilde{\mP} = \tilde{\mQ}(\tilde{\mR}_1 \; \; \tilde{\mR}_2).
\end{equation}
With $\vn$ defined as in \eqref{eq:pi}, we let 
\begin{equation}\label{eq:tilde_pi}
 \tilde{\vpi} := \tilde{\mP}^T\vn, 
\end{equation}
and the adapted design may be represented by
\begin{equation}\label{eq:design_adaption}
\vpi_{N+n} = \vpi_N \cup \braces{\text{the first } n \text{ elements of } \tilde{\vpi} \text{ (that are not in } \vpi_N) }.  
\end{equation}
Hence, the augmented design matrix may be expressed as 
\begin{equation}\label{eq:rrqr_update_psi}
\mPhi_{N+n} = \mPhi_c(\vpi_{N+n},\;:\;).
\end{equation}
\subsection{$D$-optimal Subspace Pursuit}\label{subsec:DSP}

We now have all of the necessary material to present the \textit{$D$-optimal subspace pursuit (DSP)} algorithm, which exploits the benefits of $D$-optimality to sequentially build experimental designs and perform PC approximation. DSP works by selecting rows of a candidate matrix with the methods discussed in Sections \ref{subseq:RRQR} and \ref{subsec:RRQR_adapt}. Assume we are allowed a fixed computational budget $N_{max}$, which is the maximum number of QoI evaluations to be used in order to approximate $\vc$. DSP first constructs an initial $D$-optimal design using the RRQR method via \eqref{eq:rrqr_psi} with subset selection. The initial design is constructed using fewer than $N_{max}$ samples from a candidate pool $\braces{\vxi_i}_{i=1}^M$. At each iteration, DSP updates the  estimated support set $\sS$, the design is sequentially augmented according to $\sS$ by \eqref{eq:rrqr_update_psi}, and one QoI evaluation is performed. Once $N_{max}$ samples have been selected and their corresponding QoIs have been computed, DSP is designed to approximate $\vc$ in exactly the same way as SP, i.e., DSP always performs SP iterations once $N = N_{max}$. Note that only once $N_{max}$ samples are placed in the experimental design is the complete measurement matrix formed. SP in contrast, takes as inputs a complete measurement matrix and the corresponding QoIs, which are computed  \emph{a priori}. The main steps of DSP are detailed in Algorithm \ref{alg:DSP}.
\begin{algorithm}[H]                      
\caption{$D$-optimal Subspace Pursuit (DSP)}          
\label{alg:DSP}                           
\small
\textbf{Input:} $K,  \mPhi_c, N_{max}$\\
\textbf{Initialization:} \\
 1) Let $N_0 =$ \text{max}$(2K,\lfloor 0.8 N_{max} \rfloor)$. \\
 2) $\mPhi_{N_0} = \mPhi_c( \vpi_{N_0}, \; : \; )$ corresp. to $\braces{\vxi_i}_{i=1}^{N_0}$ , and $\vv^0 = [ w(\vxi_1)u(\vxi_1),\ldots,w(\vxi_{N_0})u(\vxi_{N_0})]^T$. \\
3) $\sS^0 = \left\lbrace K \text{ indices corresp. to the largest in } |\cdot | \text{ entries of the vector } \mPhi_{N_0}^T\vv^0   \right\rbrace$. \\ 
4) $\vv_r^0 = \text{resid}(\vv^0,\mPhi_{N_0}(\;:\;,\sS^0)  ).$\\  

\textbf{Iteration:} At the $\ell$th iteration,  let $N$ = length$(\vv^{\ell-1})$ and perform the following steps:\\
1) $\tilde{\sS}^{\ell} = \sS^{\ell -1}\cup \lbrace K$ indices corresp. to the largest in $|\cdot |$ entries of the vector $\mPhi_{N}^T \vv_r^{\ell-1} \rbrace$.\\
2) Set $\hat{\vc} = \mPhi_N(\;:\;,\tilde{\sS}^{\ell}) ^\dagger\vv^{\ell-1}$.\\
3) $\sS^\ell = \lbrace K$ indices corresponding to the largest in $|\cdot |$ elements of $\hat{\vc} \rbrace$.\\
4) If $N< N_{max}$, then $\mPhi_{N+1} = \mPhi_c(\vpi_{N+1},\; : \;)$, and $\vv^{\ell} = [\vv^{\ell-1} \; ; \; w(\vxi_{N+1})u( \vxi_{N+1} )]$. \\
5) $\vv_r^\ell = \text{resid}(\vv^\ell,\mPhi_N(\;:\;,\sS^\ell)).$\\
6) If $N = N_{max}$ and $\ell=P$, quit iterating.\\
7) If $ \| \vv_r^\ell \| > \| \vv_r^{\ell -1}\|$  and $\ell \geq N_{max} -N_0 +1$, set $\sS^\ell = \sS^{\ell-1}$ and quit iterating. \\ 

\textbf{Output:} \\
1) The approximate PC coefficients $\hat{\vc}$, satisfying $\hat{\vc}(\braces{1,\ldots,P}\setminus \sS^\ell) = \bm{0}$ and $\hat{\vc}(\sS^\ell) = \mPhi_{N_{max}}(\;:\;,\sS^\ell)^\dagger \vv^\ell$.
\end{algorithm}
As mentioned in Section \ref{sec:intro}, generally speaking, the optimal value of $K$ is not known in advance and can be estimated using a cross-validation procedure. We propose one such procedure for $K$ in Section \ref{sec:numerics}; see Algorithm \ref{alg:cross_val_K}. If the optimal value of $K$ is unknown one should employ a slightly modified version of DSP; see Algorithm \ref{alg:DSP_cross_val} in Appendix A.

 In this work, we chose the size of the initial design to be of size $N_0 =$ \text{max}$(2K,\lfloor 0.8 N_{max} \rfloor)$ as an educated guess and do not claim that it is in any way optimal.  In the context of sequential sampling, existing literature does not have a well established rule of thumb for the number of initial samples. In \cite{bernardo1992integrated}, the authors suggest that the initial design should contain at least three observations per input variable. In \cite{jones1998efficient}, it is recommended that the initial design be larger using up to ten observations per input variable. The choice of $0.8N_{max}$ is based on \cite{fajraoui2017optimal}, where it was argued that the optimal initial design size was likely to be problem dependent for design adaptive PC approximation using LAR, but the best performance is achieved with a relatively large initial experimental design. For DSP, the initial design must be at least of size $2K$, otherwise the LSA in step 2) of the iteration may not be well-posed. However, in most of the numerical examples presented in Section \ref{sec:numerics}, $N_0 = \lfloor 0.8 N_{max} \rfloor>2K$ and so the initial design consists of  $\lfloor 0.8N_{max} \rfloor$ rows of the candidate matrix. Ideally, the initial design would be relatively small to obtain initial information regarding the support estimate $\sS$, and this information would be iteratively updated. If the initial design is too small, however, the support estimate could be inaccurate, leading to poor optimization of the $D$-optimal criterion in the early iterations. Even if a support index is falsely identified, DSP, SP, and CoSaMP allow for it to be removed from $\sS$ in later iterations \cite{dai2009subspace, needell2009cosamp}. This feature is in contrast to the widely used OMP algorithm, which does not allow for support indices to be removed once they are added to the set $\sS$ \cite{tropp2007signal}.

We recommend using the subset selection method as described in Section \ref{subseq:RRQR} for constructing the initial design; see step 2) of the initialization in Algorithm \ref{alg:DSP}. Recall that subset select involves both SVD and RRQR computations, but can result in a designed measurement matrix with a smaller condition number compared to using RRQR alone \cite{seshadri2016effectively, golub2012matrix}. 

\subsection{Cross validation for K}\label{subsec:cross-validation}

In most applications, the true coefficient vector $\vc$ is not likely to be exactly sparse. It may be the case that many components of $\vc$ are close to, but not exactly zero.  We assume that QoI admits an approximately sparse PC expansion, and in that regard we don't know exactly which value of $K$ results in the most accurate PC approximation. This problem affects both SP and DSP. Assume that we have a fixed number of QoI evaluations we are allowed to perform, we propose the use of a cross-validation scheme similar to that presented in \cite[Section~3.3]{doostan2011non} to approximate the optimal value of $K$. Let $N_R,N_K, $ and $N_v \in \mathbb{N}$, then we can estimate the optimal value of $K$ via the following cross-validation procedure outlined in Algorithm \ref{alg:cross_val_K}.\footnote{In all of the numerical examples presented in Section \ref{sec:numerics}, $N_R=4$, $N_K = 10$, and $N_v = N - \lfloor 0.8 N \rfloor$. }
\begin{algorithm}[!h]                     
\caption{Cross validation to estimate the optimal value of $K$ for $N$ samples with SP}          
\label{alg:cross_val_K}                           
\small
\textbf{Input:} $\mPhi\in \mathbb{R}^{N \times P} ,\vv \in \mathbb{R}^N$

\textbf{Initialize:} Let $\vk$ be a vector of $N_K$ linearly spaced integers from $1$ to $\lfloor N/2 \rfloor$, $\ve_{SP}\in \mathbb{R}^{N_K}$, and $\ve_v \in \mathbb{R}^{N_R}$.

\texttt{FOR} $i=1,\ldots,N_K$ \texttt{DO}:

\hspace{1cm}\texttt{FOR} $r=1,\ldots,N_R$ \texttt{DO}:

\hspace{2cm} 1) Let $K = \vk(i)$.

\hspace{2cm} 2) Construct $\mPhi_v$ from $N_v$ randomly selected rows of $\mPhi$ (without replacement), and let $\vv_v$ be the corresponding QoI values from $\vv$. Let the remaining rows of $\mPhi$ be denoted $\mPhi_r$, and the corresponding remaining elements of $\vv$ be denoted $\vv_r$.

\hspace{2cm} 3) Approximate $\hat{\vc}$ using SP with $\mPhi_r$ and $\vv_r$.

\hspace{2cm} 4) Compute and save $\ve_v(r) := \| \mPhi_v \hat{\vc} - \vu_v \|_2$.

\hspace{1cm} \texttt{END FOR} 

\hspace{1cm} 5) Set $\ve_{SP}(i)  = \sum_r \ve_v(r) /N_R$.

\texttt{END FOR}

6)  Let $K = \vk(i^*)$ where $i^* = \underset{i}{\text{argmin}} \; \ve_{SP}(i)$.

\textbf{Output:} $K$
\end{algorithm}
\section{Numerical Experiments}\label{sec:numerics} 

Here we consider several numerical experiments to test the ideas presented in this work. In Section \ref{subsec:doe_testing}, a comparison between coherence-optimal sampling and standard Monte Carlo sampling is performed to judge the quality of $D$-optimal designs obtained from each strategy. In Sections \ref{subsec:msf}-\ref{subsec:ishigami},  we compare three different strategies with coherence-optimal samples in terms of their ability to construct a PC approximations of a given QoI. These three strategies are coherence-optimal sampling, coherence-optimal sampling with a $D$-optimal design, and sequential coherence-optimal sampling via DSP which we denote as Coh-Opt, D-Coh-Opt, and Seq-D-Coh-Opt, respectively. We examine four different model QoIs. In Section \ref{subsec:msf} we model a QoI with manufactured sparse functions. In Section \ref{subsec:duffing} we examine the model of a nonlinear Duffing oscillator. In Section \ref{subsec:ww} we consider a model for the wing weight of a light aircraft. Finally, in Section \ref{subsec:ishigami} we discuss the Ishigami function.

In Sections \ref{subsec:msf}-\ref{subsec:ishigami}, we begin by creating a large pool of candidate samples, evaluating the polynomial basis, and computing the corresponding weights to yield a weighted candidate design matrix $\mPhi_{c} \in \mathbb{R}^{M \times P}$. Unless otherwise stated, the candidate pool is generated using coherence-optimal sampling. The coherence-optimal sampling strategy denoted Coh-Opt involves randomly selecting $N$ rows from $\mPhi_{c}$ to form a measurement matrix then using the corresponding QoI values with SP. The strategy denoted D-Coh-Opt involves using SP with an $N$-point, $D$-optimal design, namely $\mPhi_N$ as in \eqref{eq:rrqr_psi}. The sequential coherence-optimal sampling denoted Seq-D-Coh-Opt is implemented with DSP. The RRQR factorizations are computed using the \texttt{matlab qr()} function, which ensures that the absolute value of the diagonal entries of $\mR$ are in non-increasing order.

For each PC coefficient approximation using the Coh-Opt and D-Coh-Opt strategies, Algorithm \ref{alg:cross_val_K} is performed first to estimate the value of $K$, then $\hat{\vc}$ is constructed via Algorithm \ref{alg:SP}. However, since the size of the measurement matrix can change on each iteration of DSP, a modified algorithm which performs cross-validation on $K$ at each iteration is employed. For a detailed description of this modified algorithm see Algorithm \ref{alg:DSP_cross_val} in Appendix A. It should be mentioned that the computational cost associated with cross-validation of $K$  would typically be negligible compared to the cost associated with computing expensive QoIs, and it is therefore justifiable for each of the three strategies. 

For the Duffing oscillator, the wing weight model, and the Ishigami function, we assess the accuracy of the PC approximations in terms of their ability to approximate the QoI for a large number of independent validation samples. Specifically, we compute the relative validation error as 
\begin{equation}\label{eq:relative_validation_error}
e_{rel} = \frac{\| \mPsi_{val} \hat{\vc} - \vu_{val} \|_2}{\|\vu_{val}\|_2},
\end{equation} 
where $\mPsi_{val}$ is constructed by evaluating the polynomial basis using $N_{val}$ randomly sampled \emph{validation} inputs sampled from $f(\vxi)$, $\vu_{val}$ is constructed by evaluating the QoI using the validation inputs, and $\hat{\vc}$ is approximated with a separate set of \emph{reconstruction} inputs and their corresponding QoI values. Let $N$ (in the case of DSP $N=N_{max}$) be the maximum number of samples to be used in computing $\hat{\vc}$, and $\mPhi_{recon} \in \mathbb{R}^{2M \times P}$ be a weighted measurement matrix whose reconstruction inputs are sampled via coherence-optimal sampling according to \eqref{eq:Y_distribution}. The candidate design matrix is chosen by randomly selecting $M$ rows from $\mPhi_{recon}$ without replacement. The reason for constructing the candidate matrix in this manner is that we wish to repeat the PC coefficient approximation for each value of $N$, $R$ times say, for each of the three sampling strategies. Changing the candidate matrix on each repetition is necessary otherwise the designs constructed by the D-Coh-Opt and Seq-D-Coh-Opt strategies will not change. After $R$ repetitions, the mean and standard deviation of the $R$ values of $e_{rel}$ are computed. In all of the numerical results presented, $M = 10P$ and the number of validation samples is $N_{val} = 20,000$.
\subsection{The effects of sampling on $D$-optimal designs}\label{subsec:doe_testing}
As mentioned in Section \ref{subsec:doe_theory}, we expect that a candidate pool of coherence-optimal samples will produce designs with larger values of $\phi_D$ compared to a candidate pool that is constructed via standard Monte Carlo sampling. In this experiment, $D$-optimal designs are computed via \eqref{eq:rrqr_psi} with subset selection and we consider values of $\vert \tilde{\mM} \vert ^{1/P}$, where $\tilde{\mM} := \mM/ \| \mM \|_F$, and $\mM$ is defined as in \eqref{eq:weighted_information_matrix}. Note that $\vert \tilde{\mM}\vert^{1/P}$ is similar to the $D$-optimality criterion, and that larger values of $\vert \tilde{\mM} \vert ^{1/P}$ correspond to better $D$-optimal designs. The division by the Frobenius norm is to normalize so that each information matrix has the same average singular value and values of $\vert \tilde{\mM} \vert ^{1/P}$ can be compared fairly regardless of the sampling method and weights used to generate the candidate pool. 

This experiment involves constructing 1000 $D$-optimal designs comprising of $N = 250$ samples from candidate pools of $M=1000$ samples. For each design, new candidate pools are constructed independently and the value of $\vert \tilde{\mM} \vert ^{1/P}$ is recorded.  In this manner, $\vert \tilde{\mM} \vert ^{1/P}$ is considered as a random variable. For a fixed value of $N$ and $M$, comparing the empirical cumulative distribution functions (CDFs) of $\vert \tilde{\mM} \vert ^{1/P}$ provides qualitative information regarding each sampling strategy's impact on the $D$-optimal designs constructed. Specifically, given two sampling strategies MC (Monte Carlo) and Coh-Opt (coherence-optimal), we say that Coh-Opt \emph{first-order stochastically dominates} MC if $F_{MC} (\vert \tilde{\mM} \vert ^{1/P}) \leq F_{Coh-Opt} (\vert \tilde{\mM} \vert ^{1/P})$ for all $\vert \tilde{\mM} \vert ^{1/P}$ with strict inequality at some value of $\vert \tilde{\mM} \vert ^{1/P}$, where $F(\cdot)$ denotes a CDF \cite{hadar1969rules}. 

\begin{figure}[H]
\centering{
\subfloat[]{%
\includegraphics[width=0.48\textwidth]{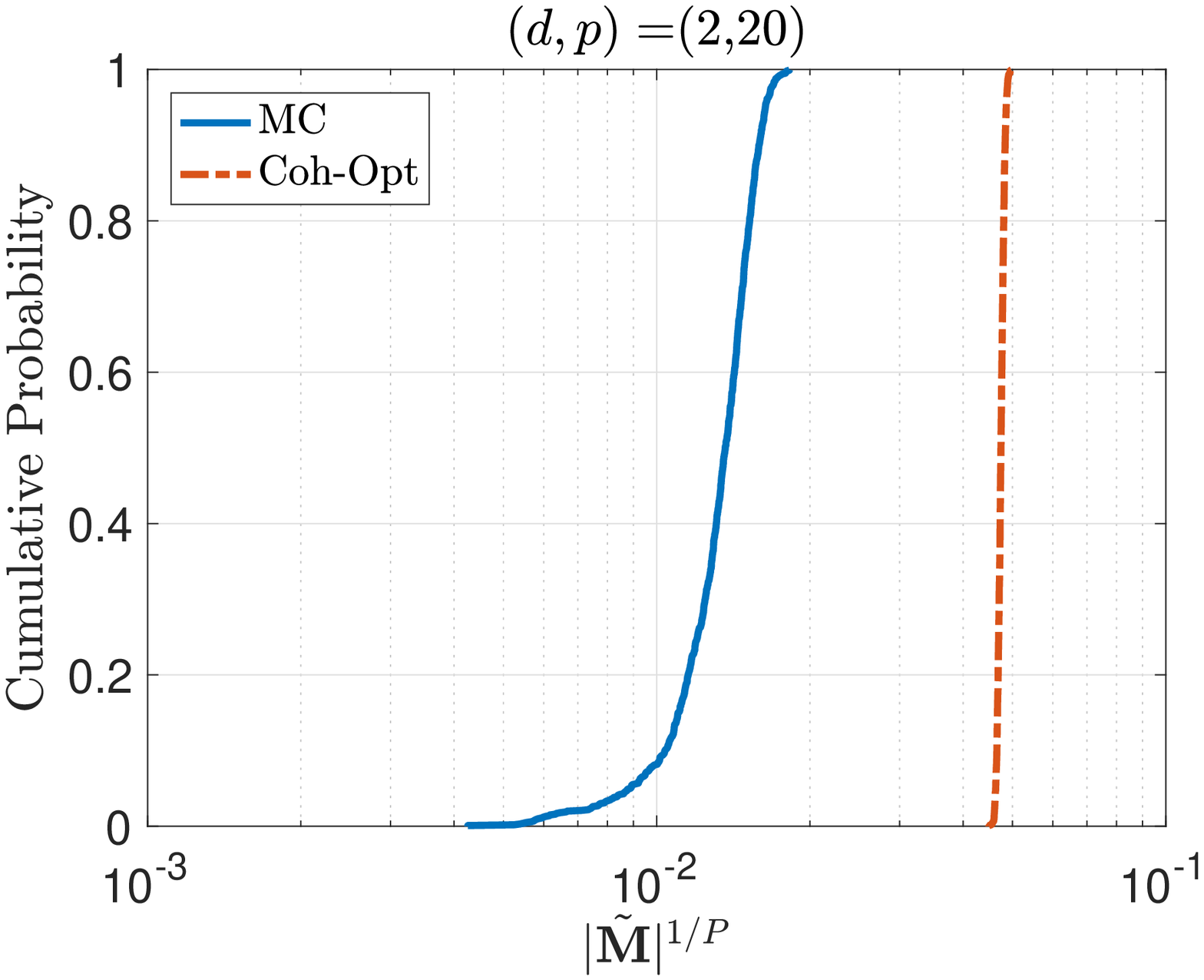}
\label{subfig:doe_test_p20_d2_L}
}
\subfloat[]{%
\includegraphics[width=0.48\textwidth]{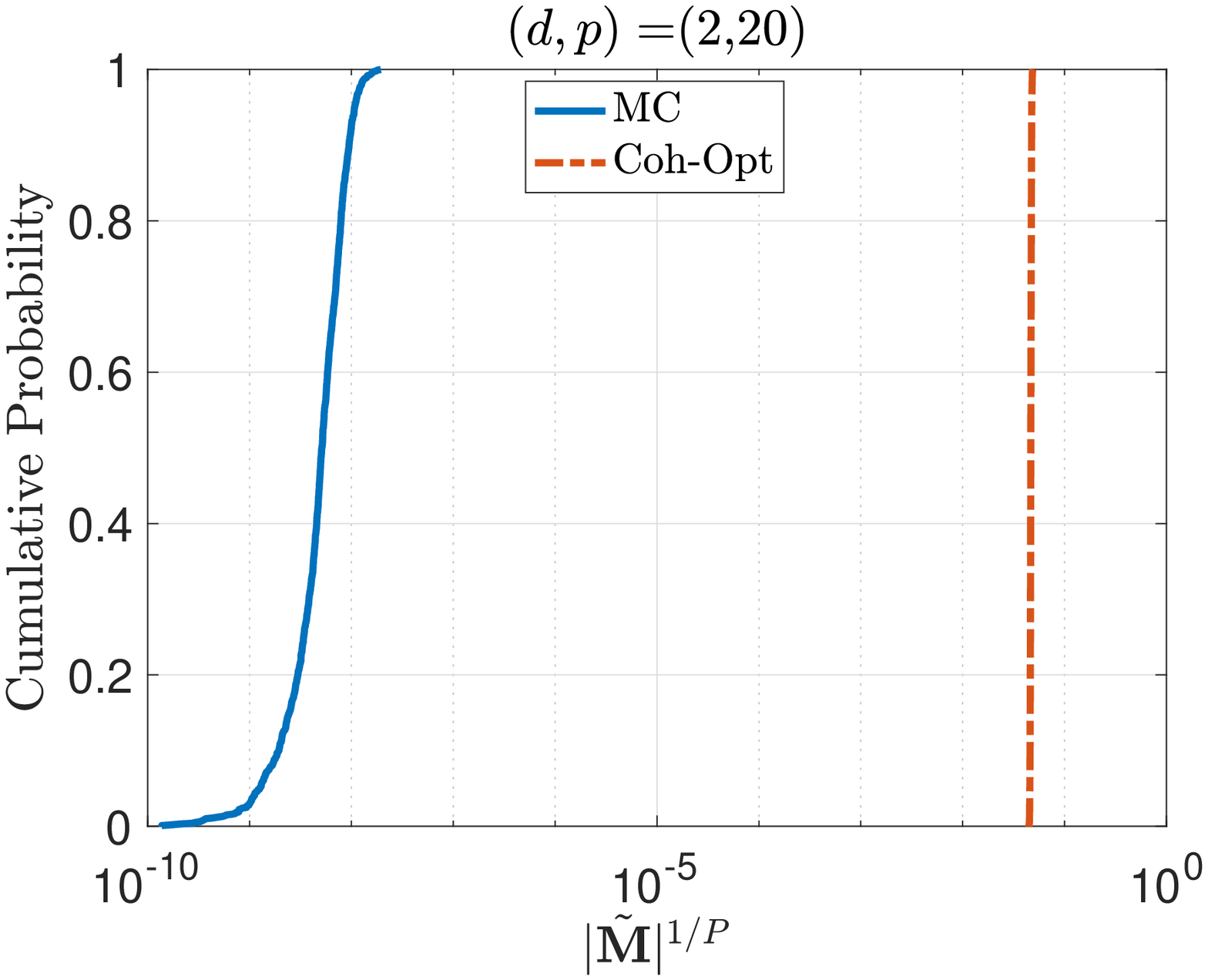}
\label{subfig:doe_test_p20_d2_h}
}
}	  
\caption{Empirical CDFs of $\vert \tilde{\mM} \vert ^{1/P}$ computed for 1000 experimental designs for $(d,p)=(2,20)$ with Legendre polynomials in \protect\subref{subfig:doe_test_p20_d2_L}, and Hermite polynomials in \protect\subref{subfig:doe_test_p20_d2_h}.}
\label{fig:doe_test_p20_d2}
\end{figure}

\begin{figure}[H]
\centering{
\subfloat[]{%
\includegraphics[width=0.48\textwidth]{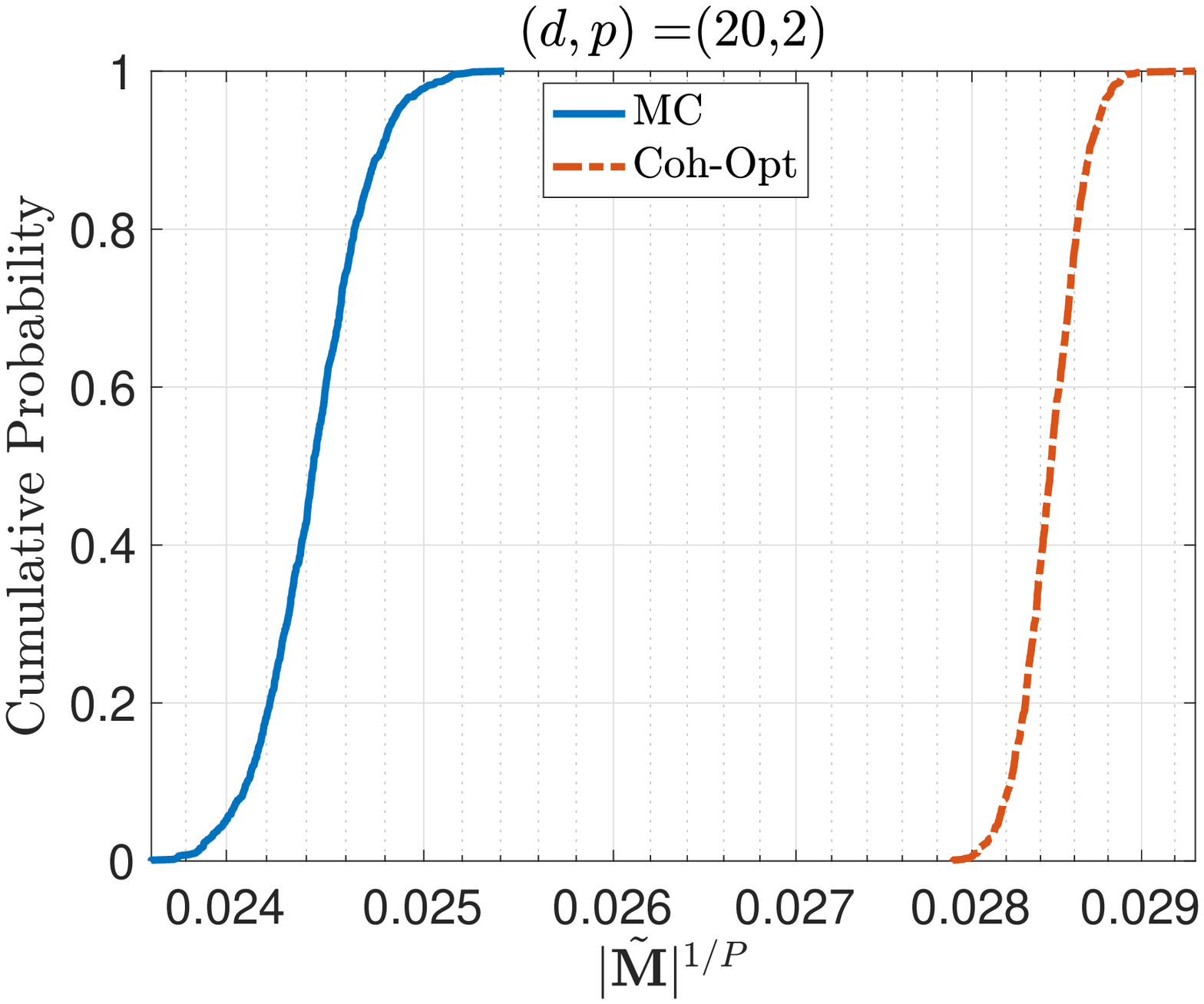}
\label{subfig:doe_test_p2_d20_L}
}
\subfloat[]{%
\includegraphics[width=0.48\textwidth]{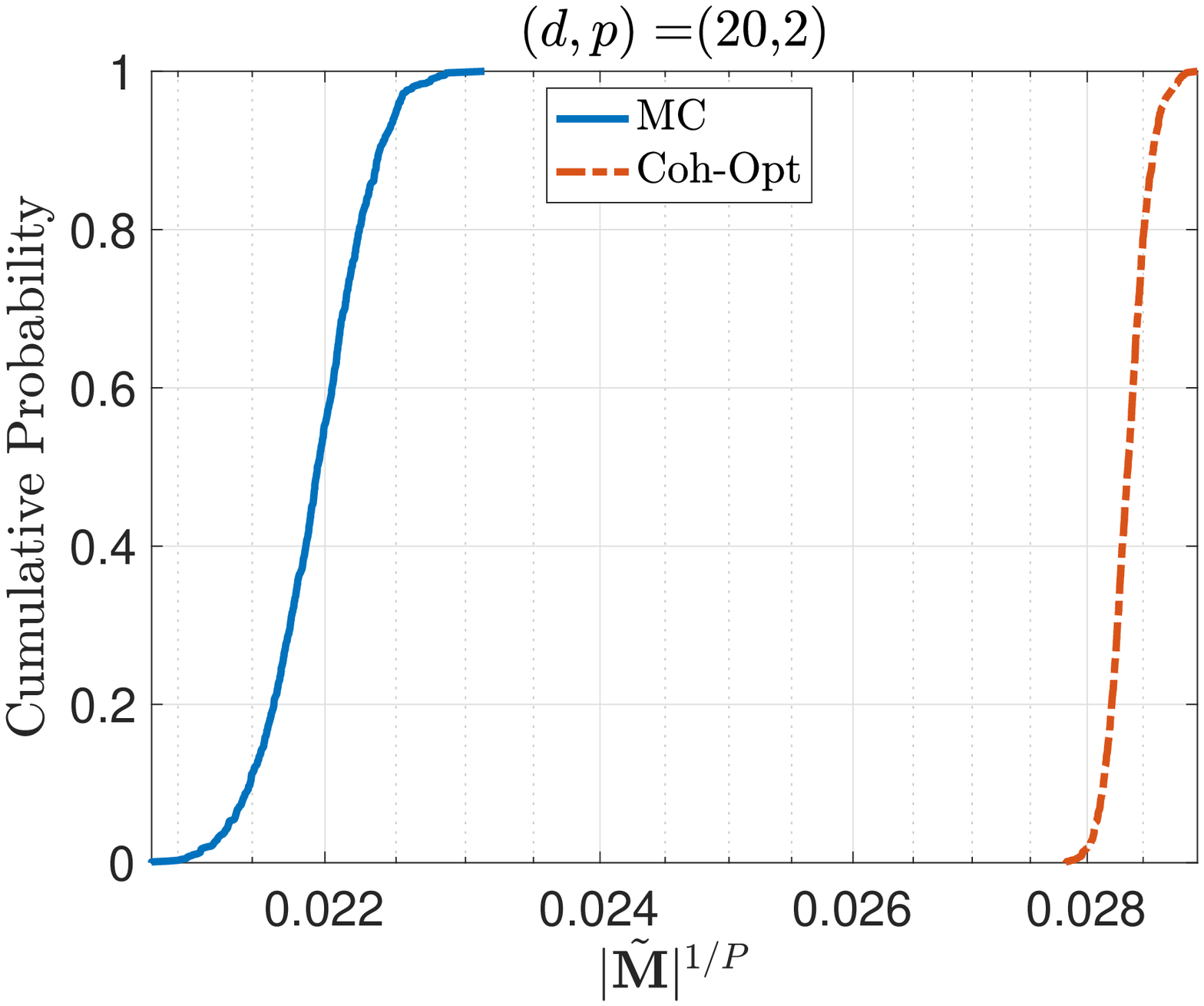}
\label{subfig:doe_test_p2_d20_h}
}
}	  
\caption{Empirical CDFs of $\vert \tilde{\mM} \vert ^{1/P}$ computed for 1000 experimental designs for $(d,p)=(20,2)$ with Legendre polynomials in \protect\subref{subfig:doe_test_p2_d20_L}, and Hermite polynomials in \protect\subref{subfig:doe_test_p2_d20_h}.}
\label{fig:doe_test_p2_d20}
\end{figure}

We consider cases for $(d,p) = (2,20)$ and $(d,p) = (20,2)$ which correspond to $P=231$ basis functions for both Legendre and Hermite polynomials. Figure \ref{fig:doe_test_p20_d2} shows the empirical CDFs of $\vert \tilde{\mM} \vert ^{1/P}$ computed using standard Monte Carlo sampling and coherence-optimal sampling for $(d,p)=(2,20)$ with Legendre and Hermite polynomials.
\noindent We see that in this case, coherence-optimal sampling produces significantly better experimental designs than standard Monte Carlo sampling. This claim is evident due to the position of the CDFs, for which coherence-optimal sampling produces larger values of $\vert \tilde{\mM} \vert ^{1/P}$ with higher probability than standard Monte Carlo sampling. Figure \ref{fig:doe_test_p2_d20} shows the empirical CDFs of $\vert \tilde{\mM} \vert ^{1/P}$ computed using standard Monte Carlo sampling and coherence-optimal sampling for $(d,p)=(20,2)$ with Legendre and Hermite polynomials.
\noindent The results for $d>p$ indicate that coherence-optimal sampling provides slightly better $D$-optimal designs. The results of this numerical experiment are consistent with those described in Section \ref{subsec:msf}, where we see that coherence-optimal sampling is particularly advantageous for the case of $d<p$, and in the case of Hermite polynomials standard Monte Carlo sampling performs significantly worse than coherence-optimal sampling. 
The greatest improvement is seen for Hermite polynomials when $d<p$.  

\subsection{Manufactured sparse PC expansions}\label{subsec:msf}

In this problem, we consider PC expansions where the vector of exact coefficients $\vc$ is manufactured. 
In our examples of manufactured sparse functions, the basis functions $\psi_k$ are multivariate Hermite polynomials, and the vector of exact coefficients $\vc$ has sparsity $s = 60$. The $s$ nonzero coefficient indices are drawn uniformly without replacement from the set $\{1,\ldots,P \}$ and their values are independent identically distributed (i.i.d.) standard normal random variables. For any given $N$,  let $M = 10P$ be the size of the candidate pool, and $R=240$. Let $\mPhi_c$ be the weighted candidate matrix corresponding to the pool of candidate samples $\{ \vxi_i \}_{i=1}^{M}$ which are drawn according to either the standard Monte Carlo or coherence-optimal sampling schemes described in Section \ref{sec:sampling}. The weighted QoIs are constructed as
\begin{equation}\label{eq:manufactured}
\vv = \mW\vu = \mW(\mPsi_c \vc+ \veps),
\end{equation}
where $\epsilon_i = \alpha |\mPsi_c(i,\;:\;)\vc | x_i$, $\mPsi_c(i,\;:\;)$ is the $i$th row of $\mPsi_c$, and $x_i \overset{\text{i.i.d.}}{\sim}N(0,1)$. In our examples, the noise level $\alpha = 0.03$.

Cross validation on $K$ is performed, even if $s$ is known in advance as it is in these examples. If one were to set $K=s$, then only in the ideal scenario -- where the support set estimate $\sS$ is perfect -- would the SP or DSP algorithms be able to approximate each of the $s$ non-zero coefficients. The reason we should allow $K>s$ is due to the fact that the QoI values, are corrupted with noise. This noise can lead to the inclusion of erroneous indices in $\sS$, and letting $K>s$ may allow for the correct $s$ non-zero coefficients to be estimated even if erroneous indices are included in $\sS$.

Given the manufactured system from \eqref{eq:manufactured}, we obtain an approximation of the exact coefficients using the  Coh-Opt, D-Coh-Opt, and Seq-D-Coh-Opt strategies. This process is repeated $240$ times for each value of $N$, i.e., $R=240$. For each repetition, we manufacture a new reference vector of coefficients $\vc$ and compute the QoIs corresponding to the candidate pool, then we compute the relative error of the PC approximations as $\|\vc -\hat{\vc}\|_2/ \|\vc \|_2$.  An \emph{oracle solution} is also computed for each repetition to gauge the performance of each sampling strategy. This oracle solution is constructed by solving an over-determined LSA using an $N$-point, $D$-optimal design with respect to the exact support set $\sS$ of $\vc$, and it represents a limit of the solution accuracy.  After $R$ repetitions are complete, we compute the mean and sample standard deviation of the relative errors.

\begin{figure}[H]    
\centering{
\subfloat[]{%
\includegraphics[width=0.49\textwidth]{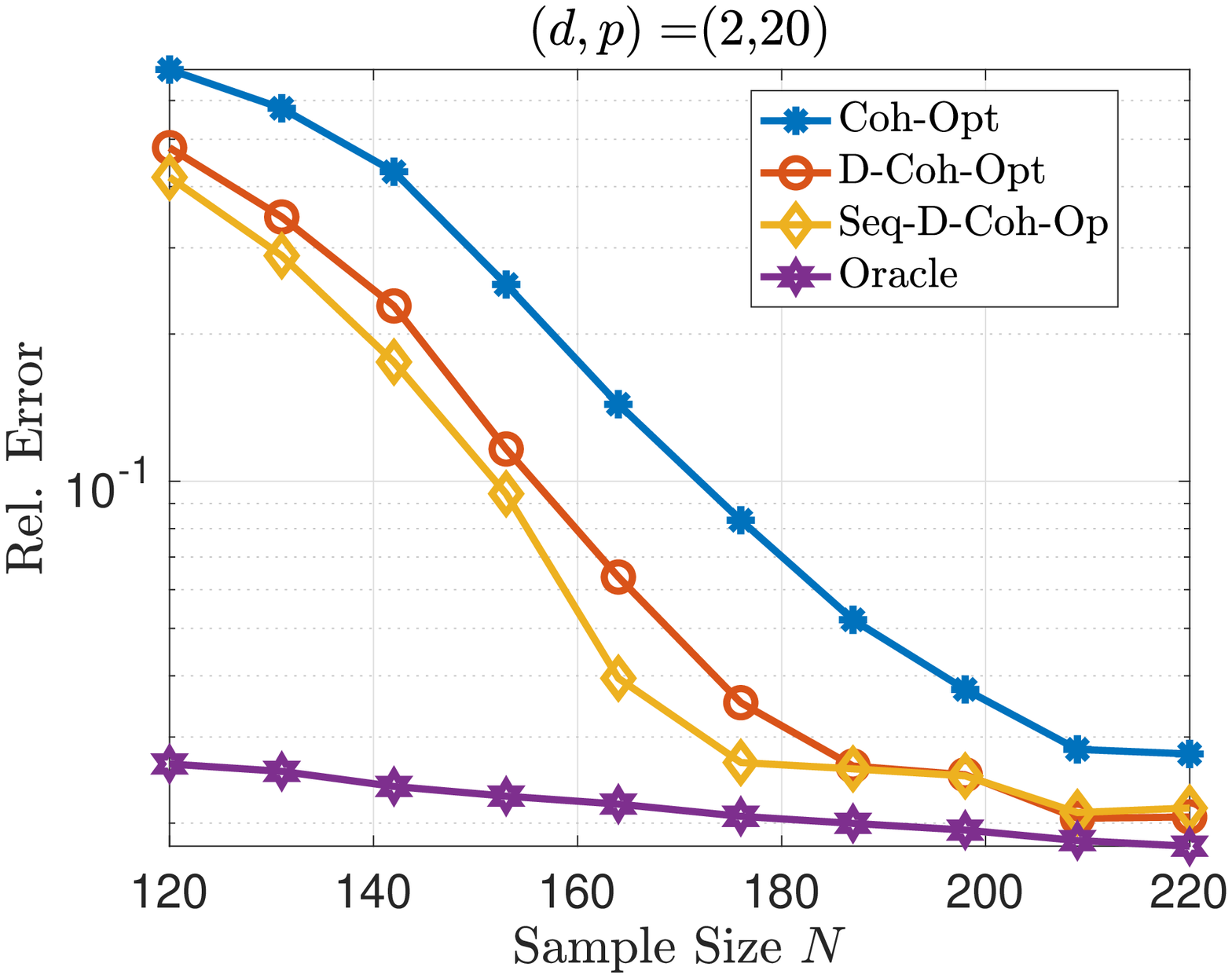}
\label{subfig:MFS_rel_err_p20_d2}
}
\subfloat[]{%
\includegraphics[width=0.49\textwidth]{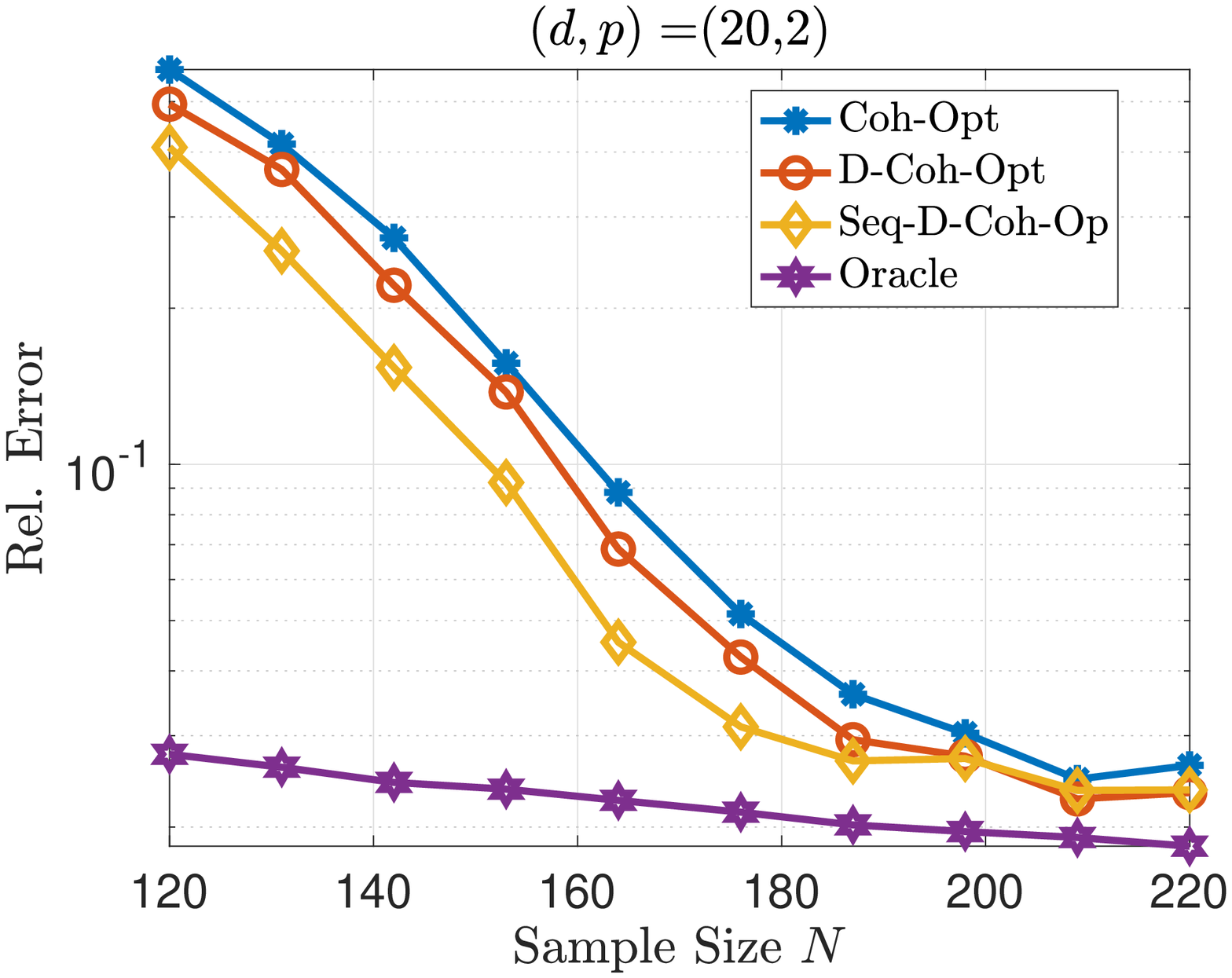}
\label{subfig:MFS_rel_err_p2_d20}
}
}	  
\caption{Mean of the relative error in approximating the exact coefficient vector $\vc$ for manufactured sparse PC expansions with sparsity $s=60$, and $(d,p)=(2,20)$ in \protect\subref{subfig:MFS_rel_err_p20_d2} and $(d,p)=(20,2)$ in \protect\subref{subfig:MFS_rel_err_p2_d20}.}
\label{fig:MFS_rel_err}
\end{figure}

\begin{figure}[H]    
\centering{
\subfloat[]{%
\includegraphics[width=0.49\textwidth]{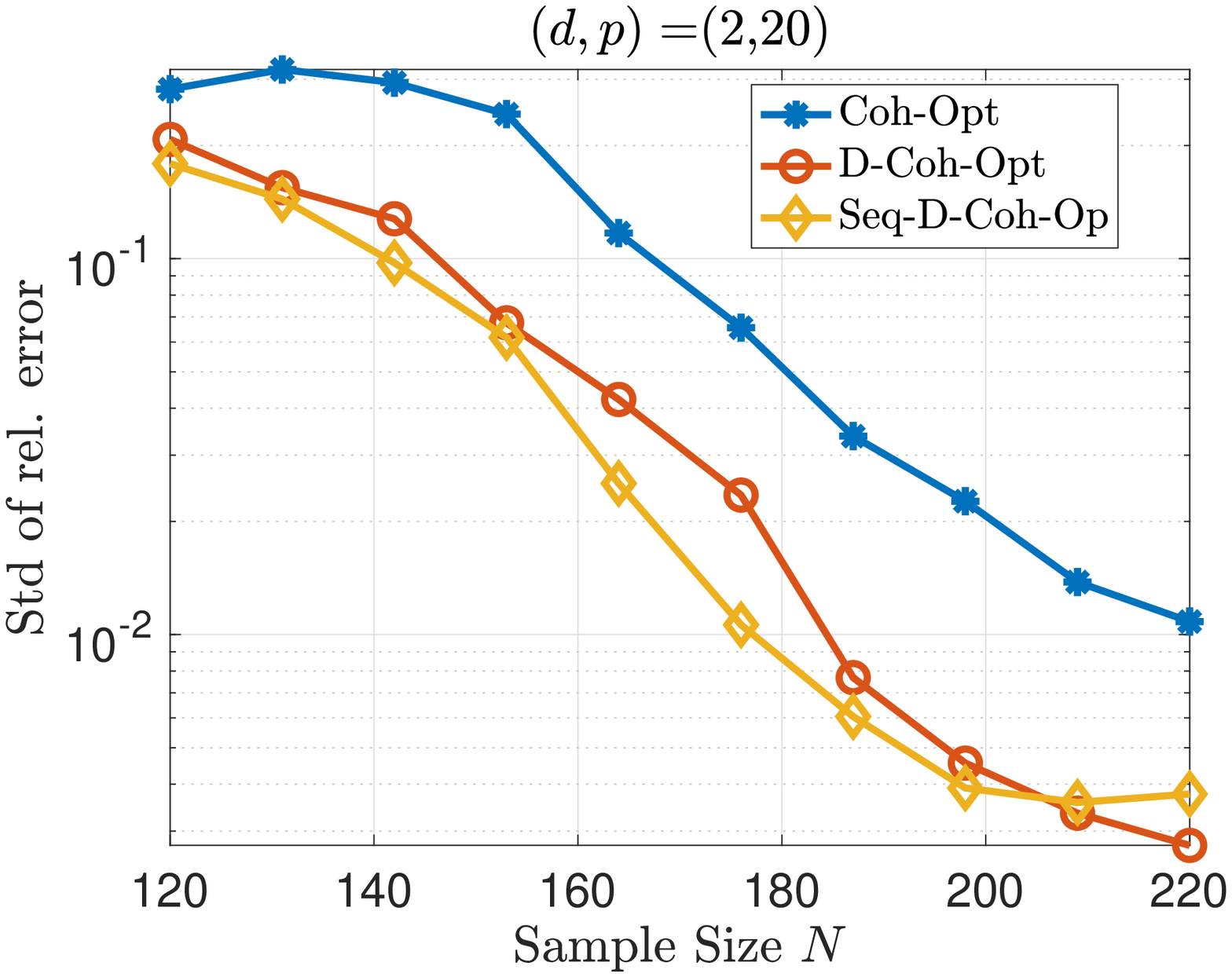}
\label{subfig:MFS_std_p20_d2}
}
\subfloat[]{%
\includegraphics[width=0.49\textwidth]{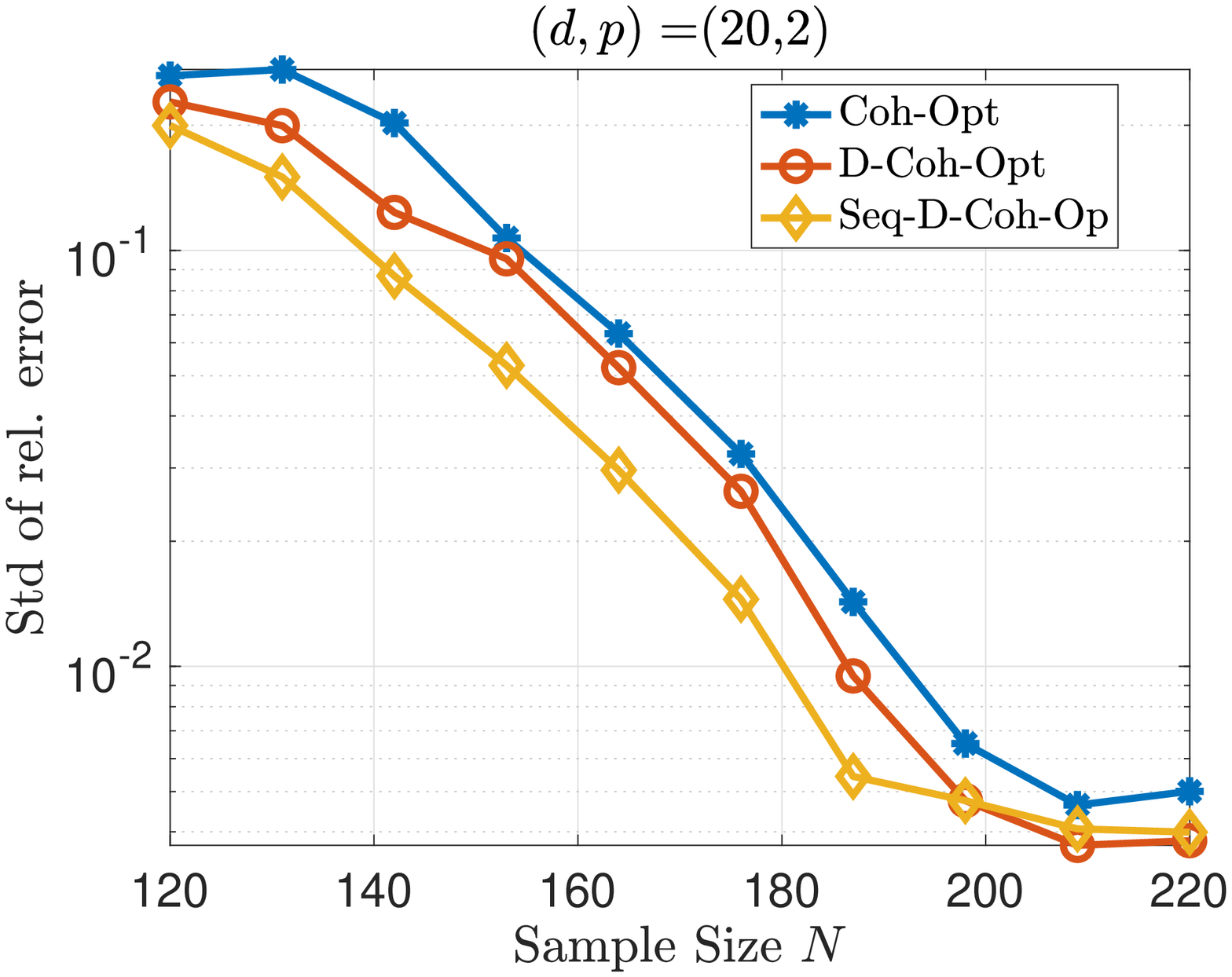}
\label{subfig:MFS_std_p2_d20}
}
}	  
\caption{Standard deviation of the relative error in approximating the exact coefficient vector $\vc$ for manufactured sparse Hermite PC expansions with sparsity $s=60$, $(d,p)=(2,20)$ in \protect\subref{subfig:MFS_std_p20_d2} and $(d,p)=(20,2)$ in \protect\subref{subfig:MFS_std_p2_d20}.}
\label{fig:MFS_std}
\end{figure}

\begin{figure}[H]    
\centering{
\subfloat[]{%
\includegraphics[width=0.49\textwidth]{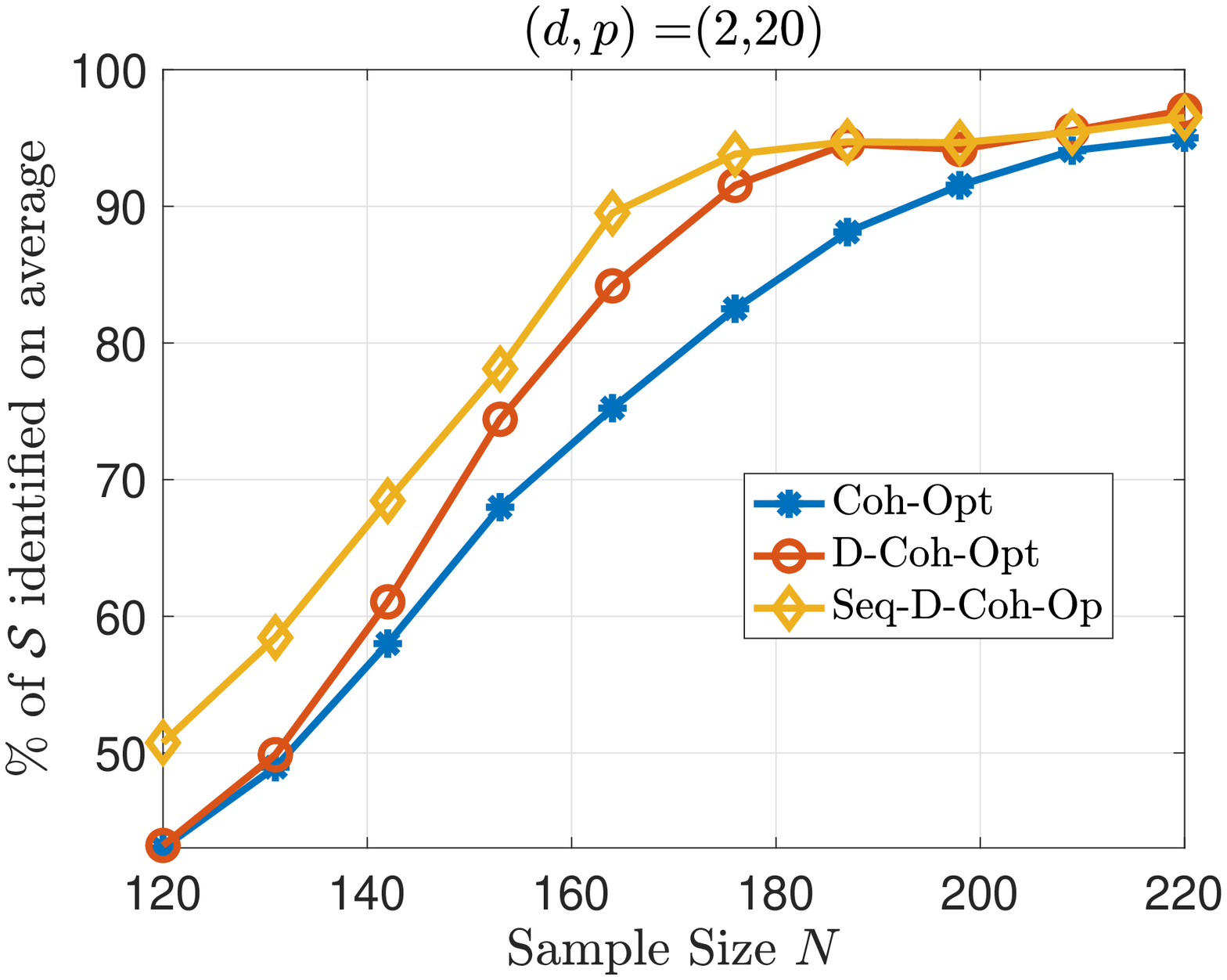}
\label{subfig:MFS_supp_p20_d2}
}
\subfloat[]{%
\includegraphics[width=0.49\textwidth]{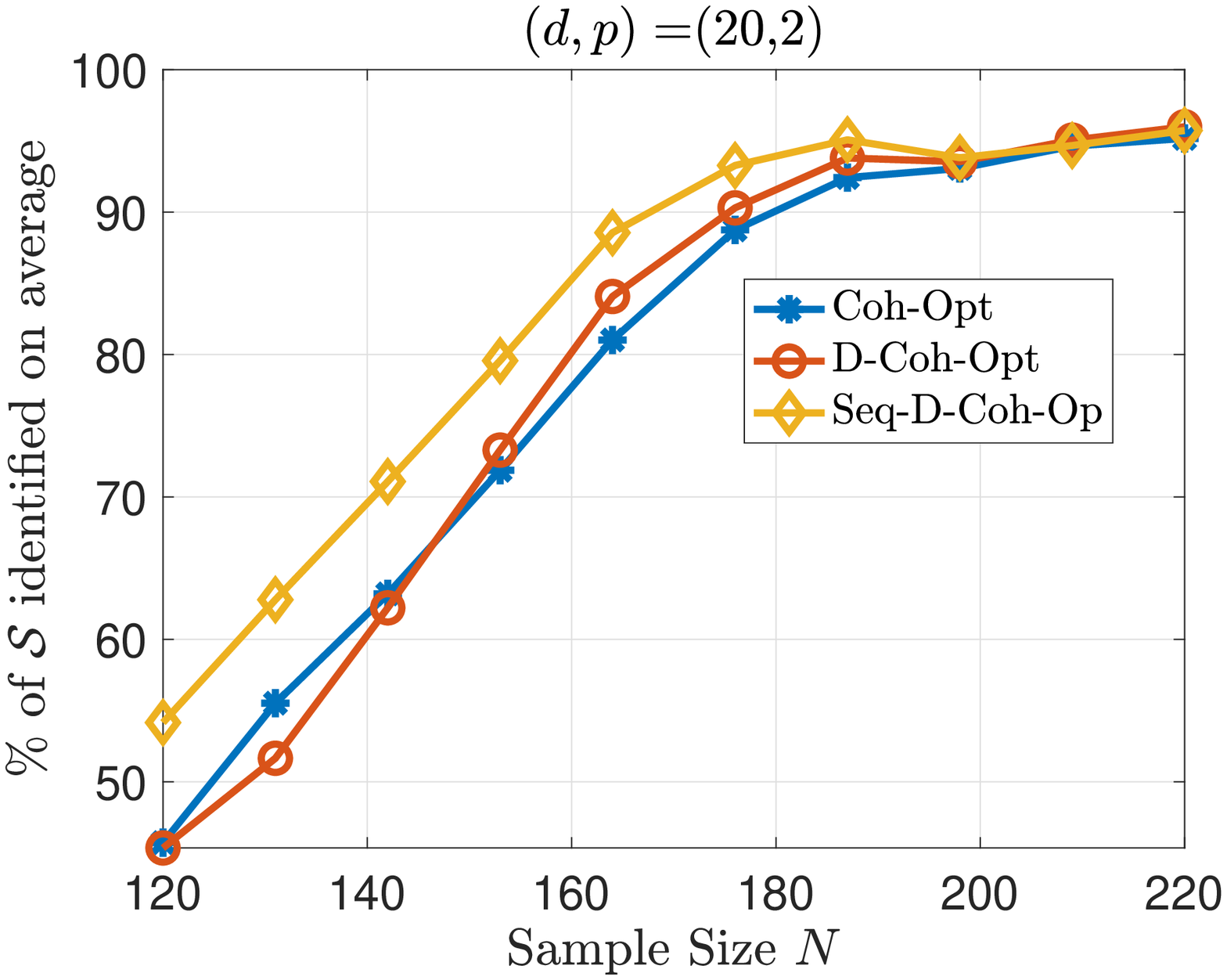}
\label{subfig:MFS_supp_p2_d20}
}
}	  
\caption{Average percentage of the support set which is correctly identified for manufactured sparse Hermite PC expansions with sparsity $s=60$, and $(d,p)=(2,20)$ in \protect\subref{subfig:MFS_supp_p20_d2} and $(d,p)=(20,2)$ in \protect\subref{subfig:MFS_supp_p2_d20}.}
\label{fig:MFS_supp}
\end{figure}

Figure \ref{fig:MFS_rel_err} depicts the mean relative error in approximating the exact coefficient vector for $(d,p)=(2,20)$ and $(d,p)=(20,2)$, both of which correspond to $P=231$ basis functions. Figure \ref{fig:MFS_std} shows the standard deviation of the relative errors, and Figure \ref{fig:MFS_supp} depicts the percentage of the support set $\sS$ which is correctly identified, on average. By each of the three metrics, Seq-D-Coh-Opt shows the best performance of the three strategies, Coh-Opt performs the worst, and D-Coh-Opt shows intermediate performance. In the low-dimension, high-order case, D-Coh-Opt and Seq-D-Coh-Opt show greater improvements in accuracy compared to Coh-Opt than they do in the high-dimension, low-order case. This result is consistent with the findings of Section \ref{subsec:doe_testing}. 

\begin{figure}[H]    
\centering{
\subfloat[]{%
\includegraphics[width=0.49\textwidth]{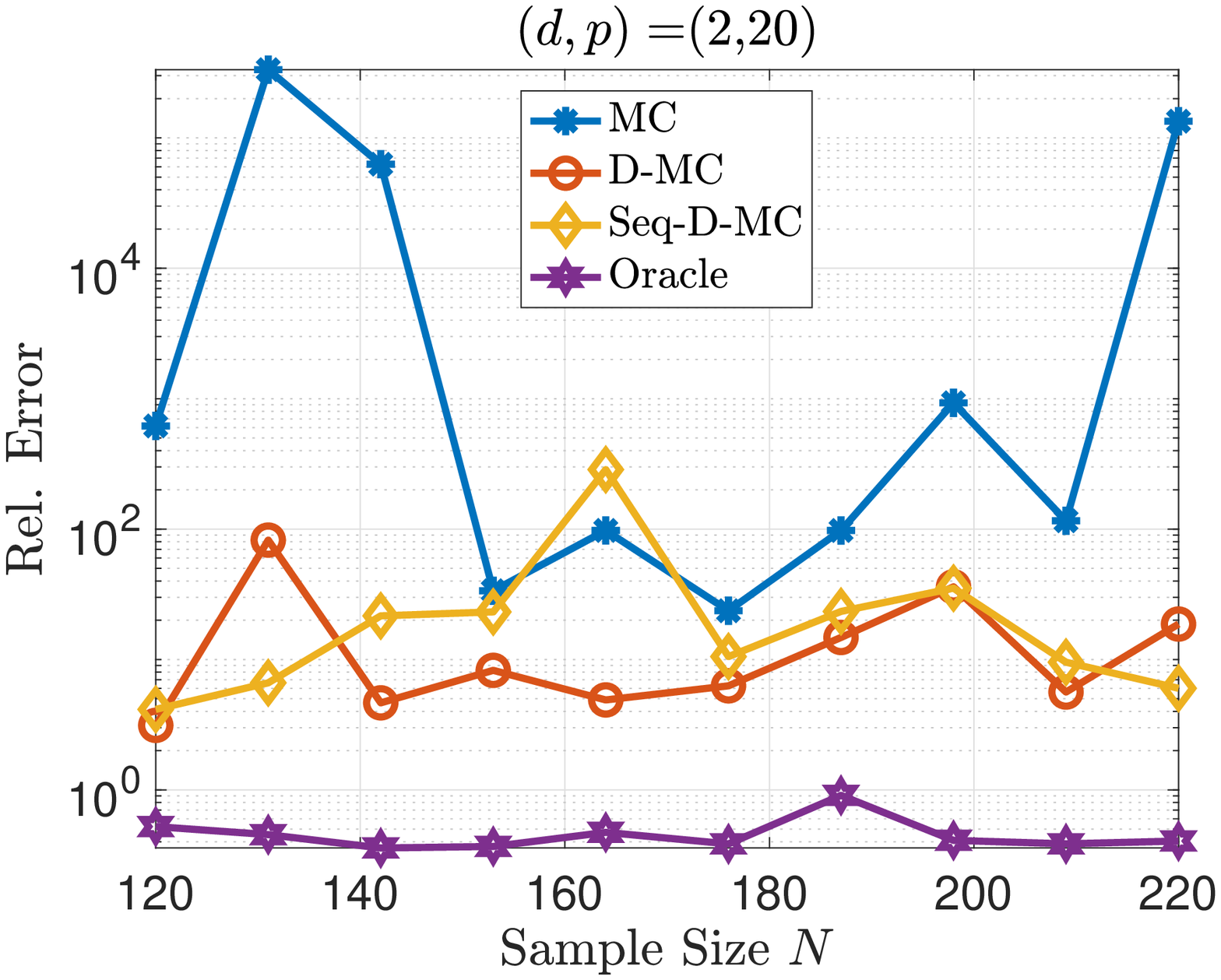}
\label{subfig:MFS_rel_err_standard}
}
\subfloat[]{%
\includegraphics[width=0.49\textwidth]{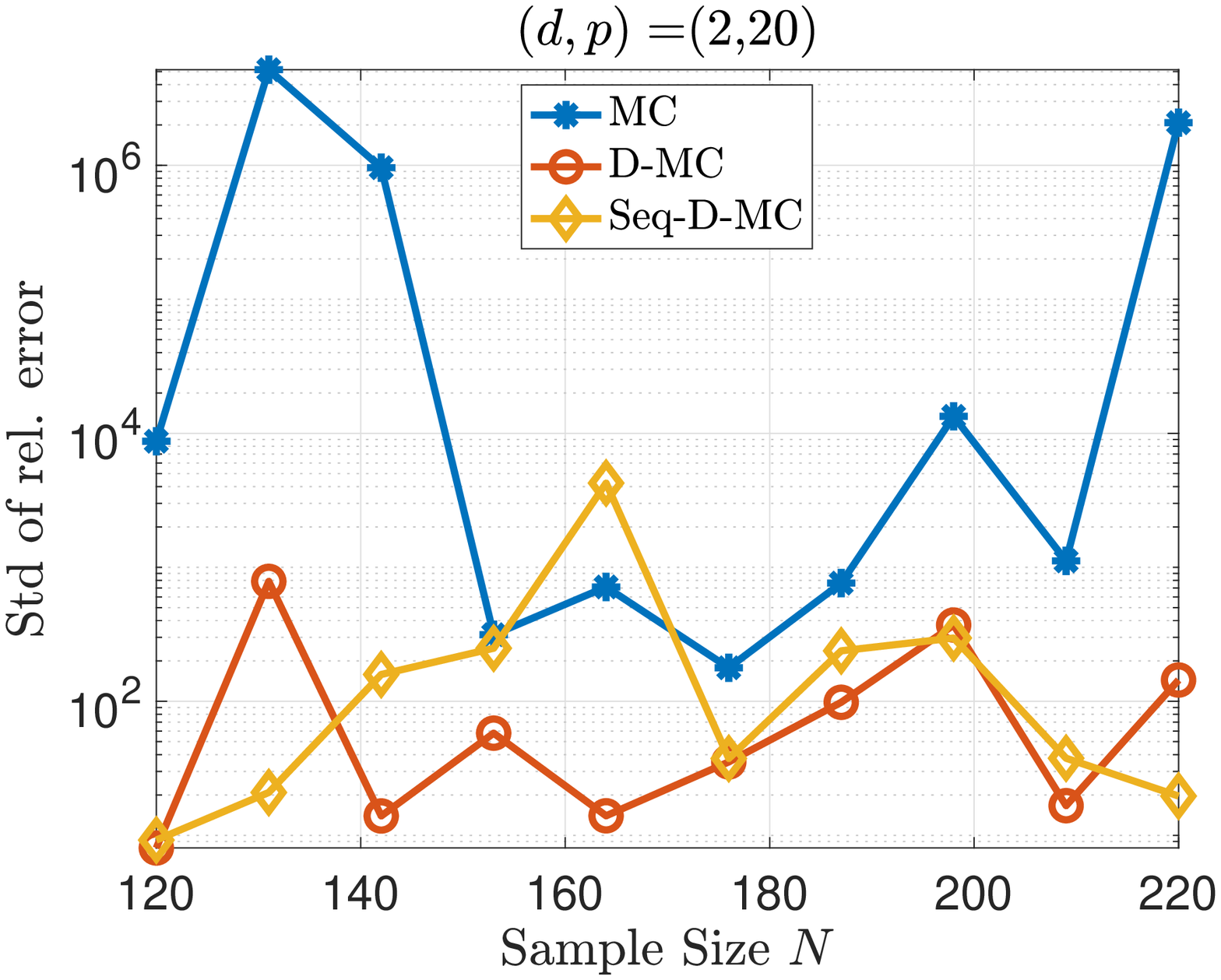}
\label{subfig:MFS_std_standard}
}\\
\subfloat[]{%
\includegraphics[width=0.49\textwidth]{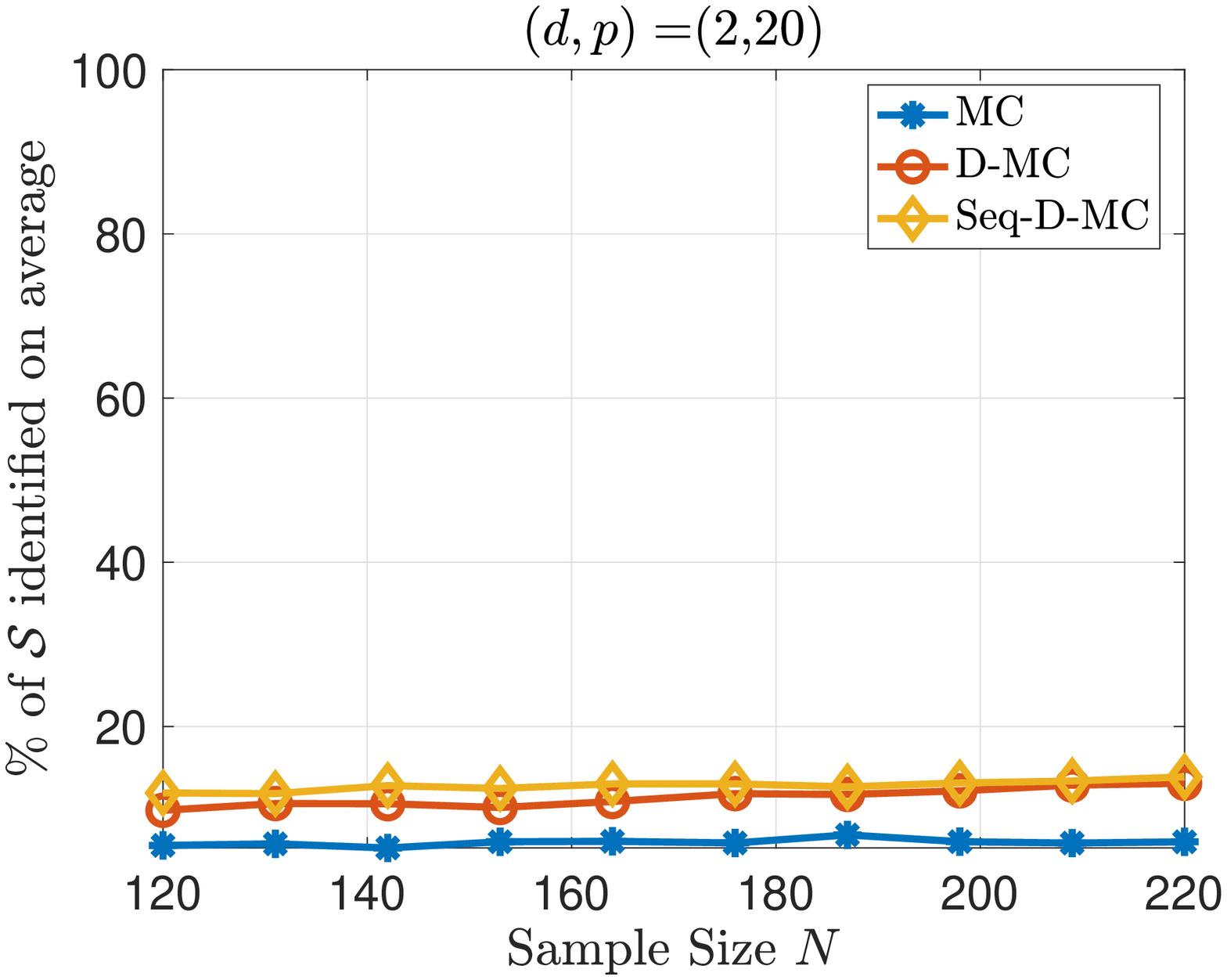}
\label{subfig:MFS_supp_standard}
}
}	  
\caption{ This example uses standard Monte Carlo sampling for $(d,p)= (2,20)$ manufactured sparse Hermite PC expansions, which results in unstable and inaccurate approximations of $\vc$. The average relative errors, standard deviations of relative errors, and percentage of the support set which is correctly identified are depicted in Figures \protect\subref{subfig:MFS_rel_err_standard}, \protect\subref{subfig:MFS_std_standard}, and \protect\subref{subfig:MFS_supp_standard}, respectively. MC denotes standard Monte Carlo sampling with SP, D-MC denotes an $N$-point $D$-optimal design of Monte Carlo samples with SP, and Seq-D-MC denotes sequential Monte Carlo sampling via DSP. }
\label{fig:MFS_standard}
\end{figure}

To demonstrate the benefits of coherence-optimal sampling, we repeat the experiment for $(d,p)=(2,20)$, this time using a candidate pool of standard Monte Carlo samples, i.e., sampling according to $f(\vxi)$ with $\mW = \mI$. The results of this experiment are presented in Figure \ref{fig:MFS_standard}, where MC denotes standard Monte Carlo sampling with SP, D-MC denotes an $N$-point, $D$-optimal design of Monte Carlo samples with SP, and Seq-D-MC denotes sequential Monte Carlo sampling via DSP. We observe that each of the three strategies fail to construct accurate, stable PC coefficient approximations when standard Monte Carlo sampling is used. This finding is not particularly surprising given the results of Section \ref{subsec:doe_testing}, which indicate the standard Monte Carlo sampling for large $p$ results in very poor experimental designs for Hermite polynomials. Further, these results are fundamentally different to those presented in Figure \ref{fig:MFS_rel_err}, where coherence-optimal sampling yields accurate and stable PC coefficient approximations. 

\subsection{A Nonlinear Duffing Oscillator}\label{subsec:duffing}

Here we consider the displacement solution $u(\vXi,t)$ of a nonlinear single-degree-of-freedom Duffing oscillator \cite{mai2015polynomial,hadigol2017least} under free vibration described by
\begin{equation}\label{eq:duffing}
\begin{aligned}
&\ddot{u}(\vXi,t)+ 2\omega_1\omega_2 \dot{u}(\vXi,t)+\omega_1^2 (u(\vXi,t)+\omega_3u^3(\vXi,t)) = 0, \\
& u (\vXi,0) = 1, \quad \dot{u}(\vXi,0)=0, \\
\end{aligned}
\end{equation}
and with uncertain input parameters $\braces{\omega_i}_{i=1}^3$ such that
\begin{equation}\label{eq:duffing_inputs}
\begin{aligned}
\omega_1 &= 2\pi(1+0.2\Xi_1),\\
\omega_2 &= 0.05(1+0.05\Xi_2),\\
\omega_3 &= -0.5(1+0.5\Xi_3),
\end{aligned}
\end{equation}
where $\Xi_i \overset{\text{i.i.d.}}{\sim}  U(-1,1)$. Our QoI is $u(\vXi,4)$.
Although this is a relatively low-dimensional problem, due to the stochastic frequency of oscillations, high-order PC expansions are required to maintain a fixed level of accuracy at large time instances $t$ \cite{hadigol2017least}.  We consider $p = 9,12$ for this problem to investigate the performance of our sampling strategies. The pair $(d,p) = (3,9)$ leads to $P=220$ PC coefficients to be approximated, whereas the pair $(d,p) = (3,12)$ leads to $P=455$ PC coefficients. For any given $N$,  let $M = 10P$ be the size of the candidate pool and $R=1000$.

\begin{figure}[H]    
\centering{
\subfloat[]{%
\includegraphics[width=0.49\textwidth]{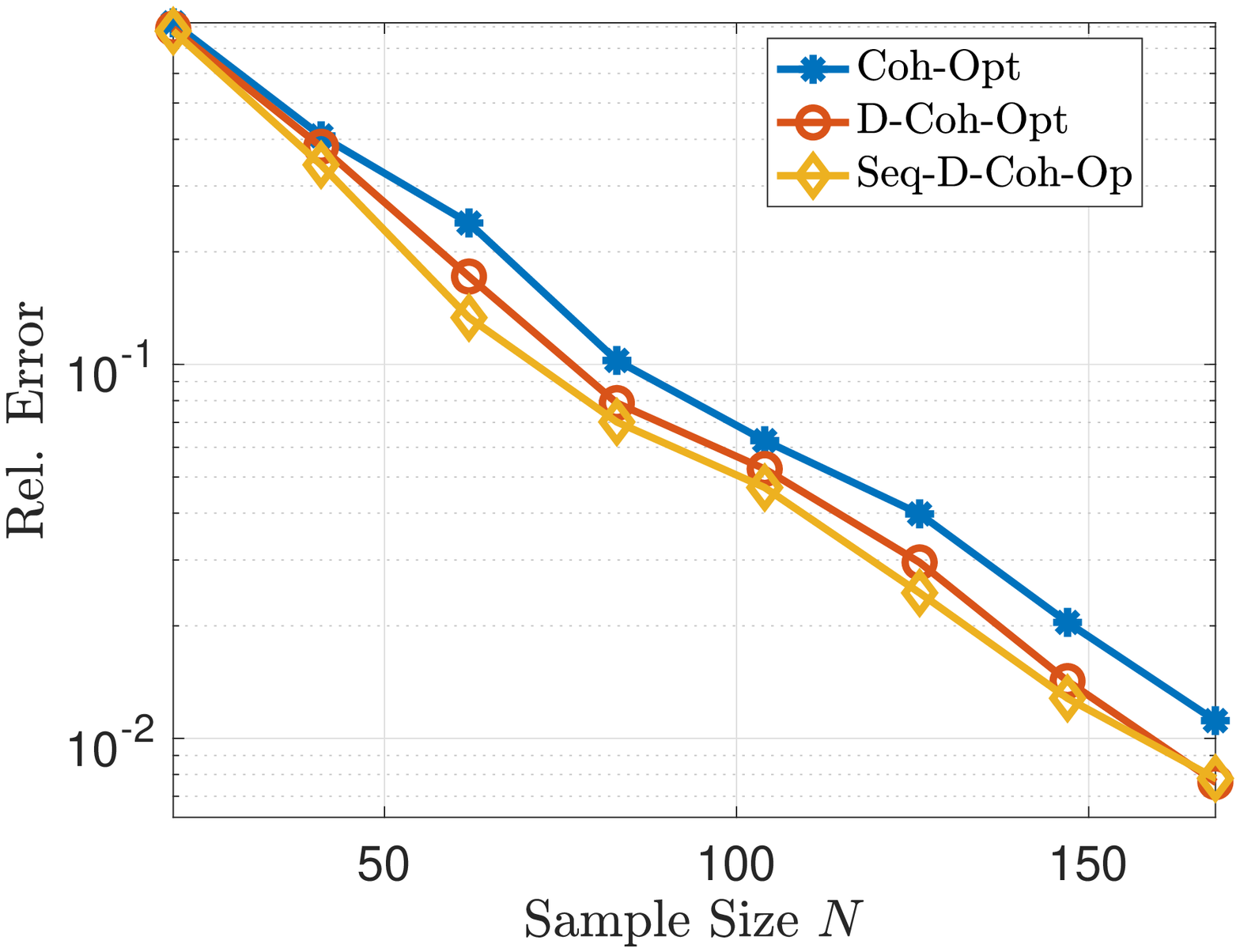}
\label{subfig:duffing_rel_err_p9}
}
\subfloat[]{%
\includegraphics[width=0.49\textwidth]{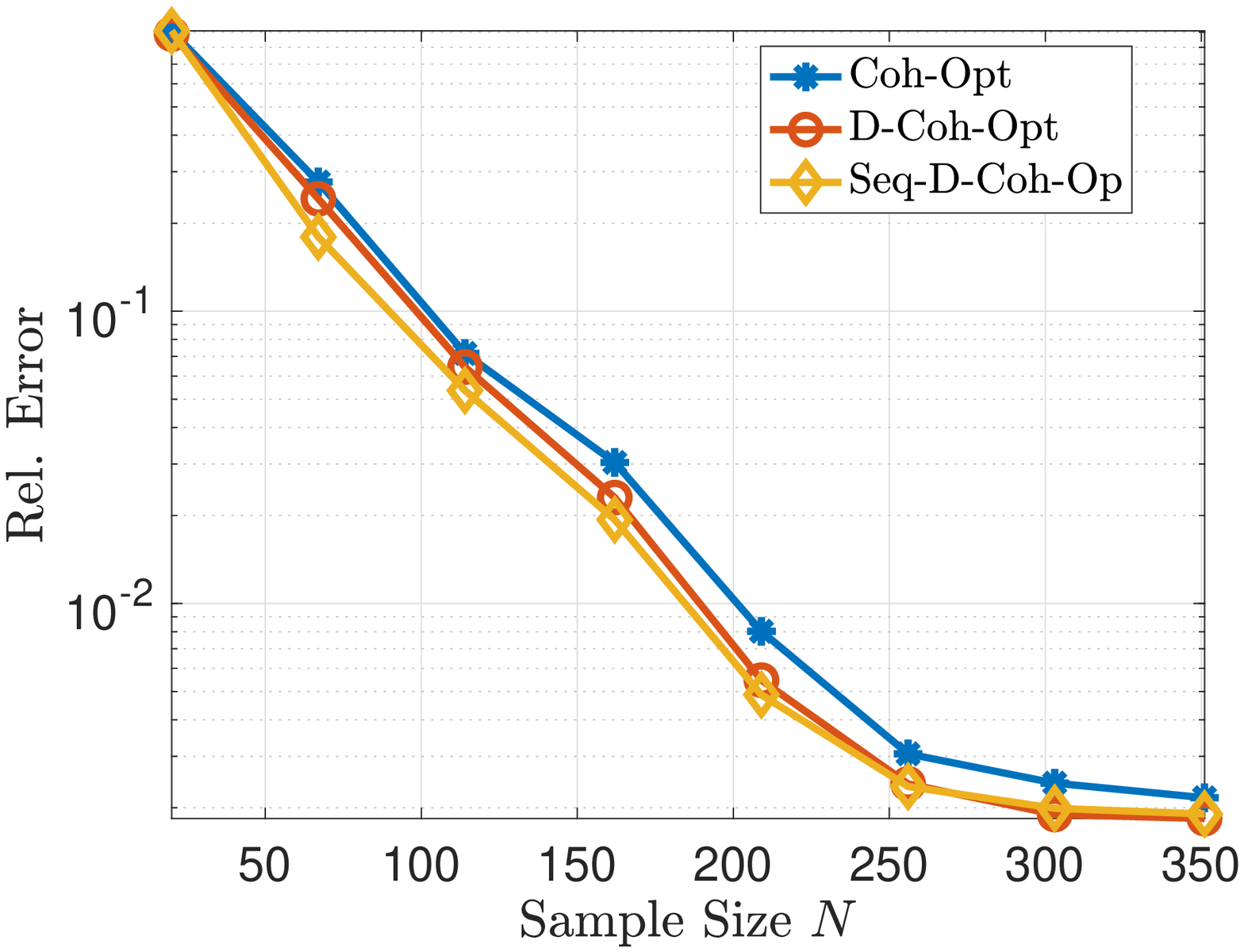}
\label{subfig:duffing_rel_err_p12}
}
}	  
\caption{Mean of the relative error in estimating the displacement $u(\vXi,4)$ with a 9th and 12th order PC expansion in \protect\subref{subfig:duffing_rel_err_p9} and \protect\subref{subfig:duffing_rel_err_p12}, respectively.}
\label{fig:duffing_rel_err}
\end{figure}

\begin{figure}[H]    
\centering{
\subfloat[]{%
\includegraphics[width=0.49\textwidth]{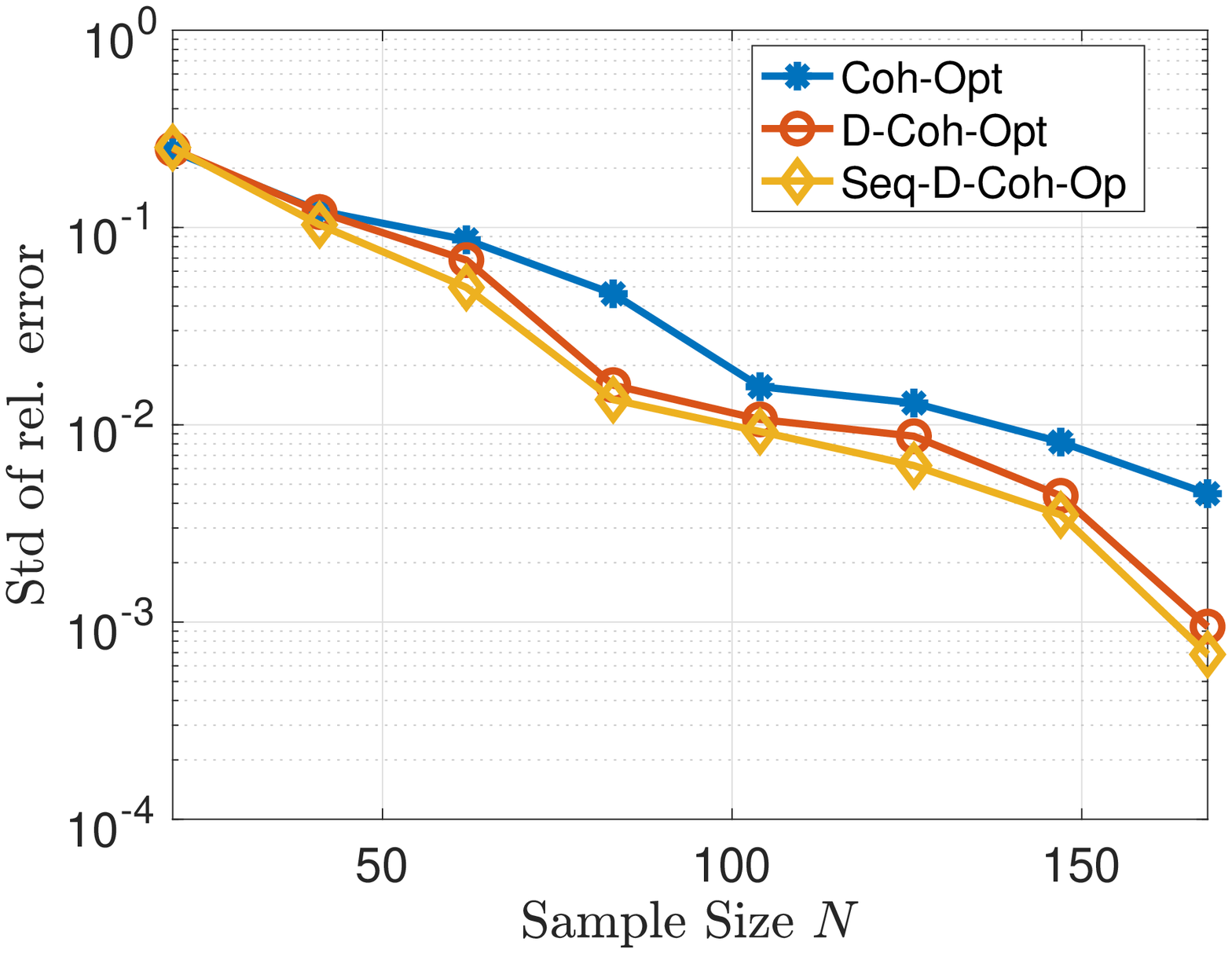}
\label{subfig:duffing_std_p9}
}
\subfloat[]{%
\includegraphics[width=0.49\textwidth]{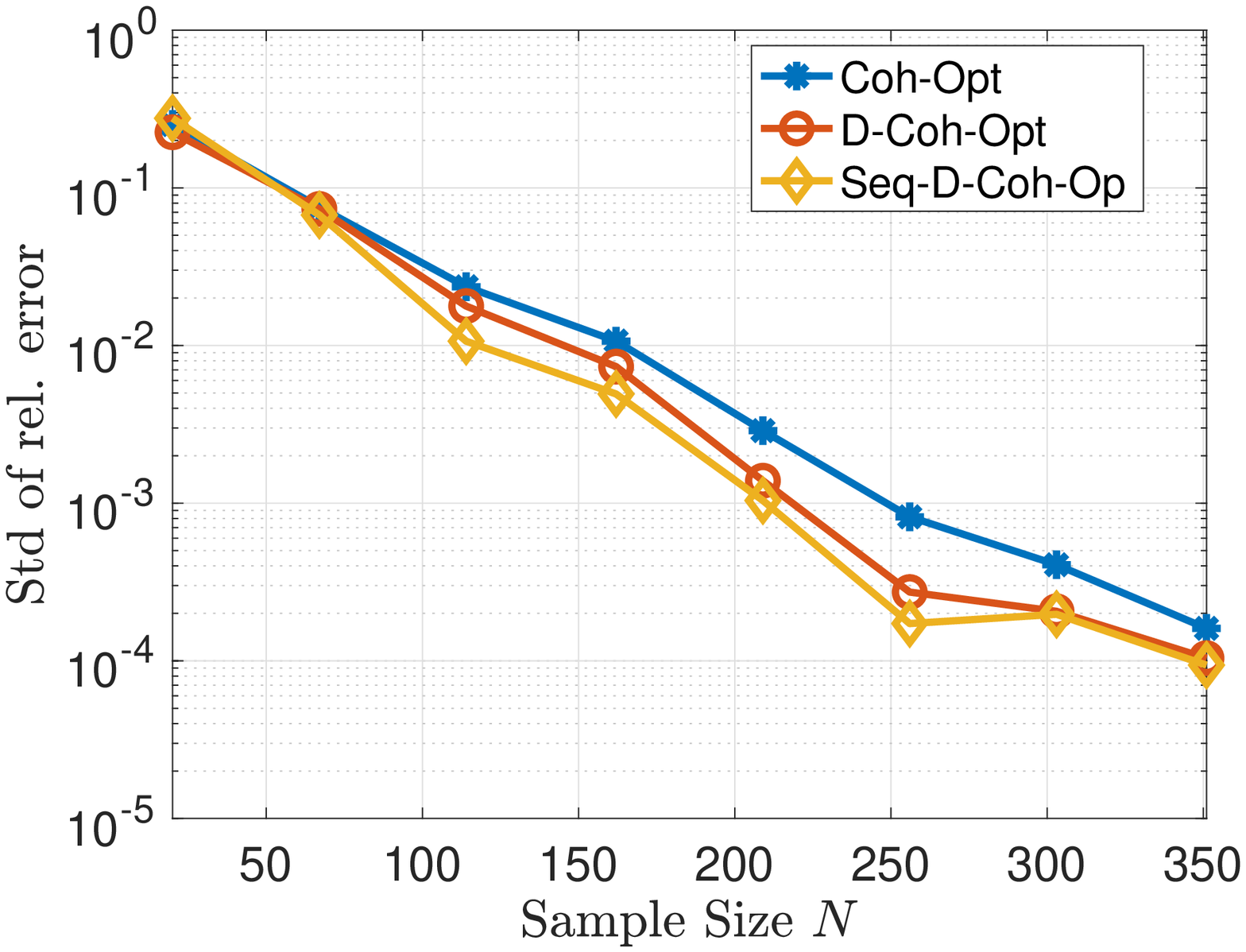}
\label{subfig:duffing_std_p12}
}
}	  
\caption{Standard deviation of the relative error in estimating the displacement $u(\vXi,4)$ with a 9th and 12th order PC expansion in \protect\subref{subfig:duffing_rel_err_p9} and \protect\subref{subfig:duffing_rel_err_p12}, respectively.}
\label{fig:duffing_std}
\end{figure}

Figures \ref{fig:duffing_rel_err} and \ref{fig:duffing_std} demonstrate the mean and standard deviation of the relative validation errors, respectively. 

We see that D-Coh-Opt and Seq-D-Coh-Opt clearly result in more accurate PC coefficient approximations, with less variability, compared to Coh-Opt. Further, these figures demonstrate that Seq-D-Coh-Opt results in slightly more accurate approximations than the non-sequential D-Coh-Opt strategy. These results are in agreement with those presented in Section \ref{subsec:msf}.  As one might expect, the PC approximations using $p=12$ achieve more accurate results than do those constructed using $p=9$. 

\subsection{The Wing Weight Function}\label{subsec:ww}

In this problem, we investigate a wing weight function which models a light aircraft wing. Details of the model and the associated \texttt{Matlab} code are available at \url{https://www.sfu.ca/~ssurjano/wingweight.html}. The wing weight model also appears in \cite{forrester2008engineering,moon2010design,moon2012two} as a test model for statistical screening. The model response is the wing's weight $u(\vXi)$, given by the nonlinear expression
\begin{equation}\label{eq:ww}
u(\vXi)  = 0.036S_w^{0.758}W_{f_w}^{0.0035}\parens{\frac{A}{\cos^2(\Lambda)}}^{0.6}q^{0.006}\lambda^{0.04}\parens{\frac{100t_c}{\cos(\Lambda)}}^{-0.3}(N_z W_{dg})^{0.49}+S_wW_p,
\end{equation}
where each of the 10 input parameters' descriptions, ranges, and units are described in Table \ref{tab:ww}. 
\begin{table}[H]
\parbox{1\linewidth}{\centering \footnotesize
\begin{tabular}{|| c | c ||}
	\hline
	\multicolumn{2}{|| c ||}{\textbf{Input Parameters}} \\
	\hline
	\pbox{\textwidth}{$S_w\in[150,200]$}&\pbox{\textwidth}{wing area (ft$^2$)}  \\ 
	\hline 
	\pbox{\textwidth}{$W_{f_w} \in [220,300]$}&\pbox{\textwidth}{weight of fuel in the wing (lb)} \\
	\hline
    \pbox{\textwidth}{$A\in[6,10]$}&\pbox{\textwidth}{aspect ratio} \\
	\hline
    \pbox{\textwidth}{$\Lambda \in [-10,10]$}&\pbox{\textwidth}{quarter-chord sweep (degrees)} \\
	\hline
	\pbox{\textwidth}{$q \in [16,45]$}&\pbox{\textwidth}{dynamic pressure at cruise (lb/ft$^2$)} \\
	\hline
    \pbox{\textwidth}{$\lambda \in [0.5,1]$}&\pbox{\textwidth}{taper ratio} \\
	\hline
    \pbox{\textwidth}{$t_c \in [0.08, 0.18]$}&\pbox{\textwidth}{aerofoil thickness to chord ratio} \\
	\hline
	\pbox{\textwidth}{$N_z \in [2.5,6]$}&\pbox{\textwidth}{ultimate load factor} \\
	\hline
	\pbox{\textwidth}{$W_{dg}\in [1700,2500]$}&\pbox{\textwidth}{flight design gross weight (lb)} \\
	\hline
	\pbox{\textwidth}{$W_p \in [0.025,0.08]$}&\pbox{\textwidth}{paint weight (lb/ft$^2$)} \\
	\hline
\end{tabular}
}
\caption{Input parameters for the wing weight function described by \eqref{eq:ww}. Each of the parameters corresponds to a $\Xi_i\overset{\text{i.i.d.}}{\sim}U(-1,1),$ $i=1,\ldots,10,$ which is shifted and scaled to be in the parameter domains defined above. } \label{tab:ww}
\end{table}

\begin{figure}[!h]    
\centering{
\subfloat[]{%
\includegraphics[width=0.49\textwidth]{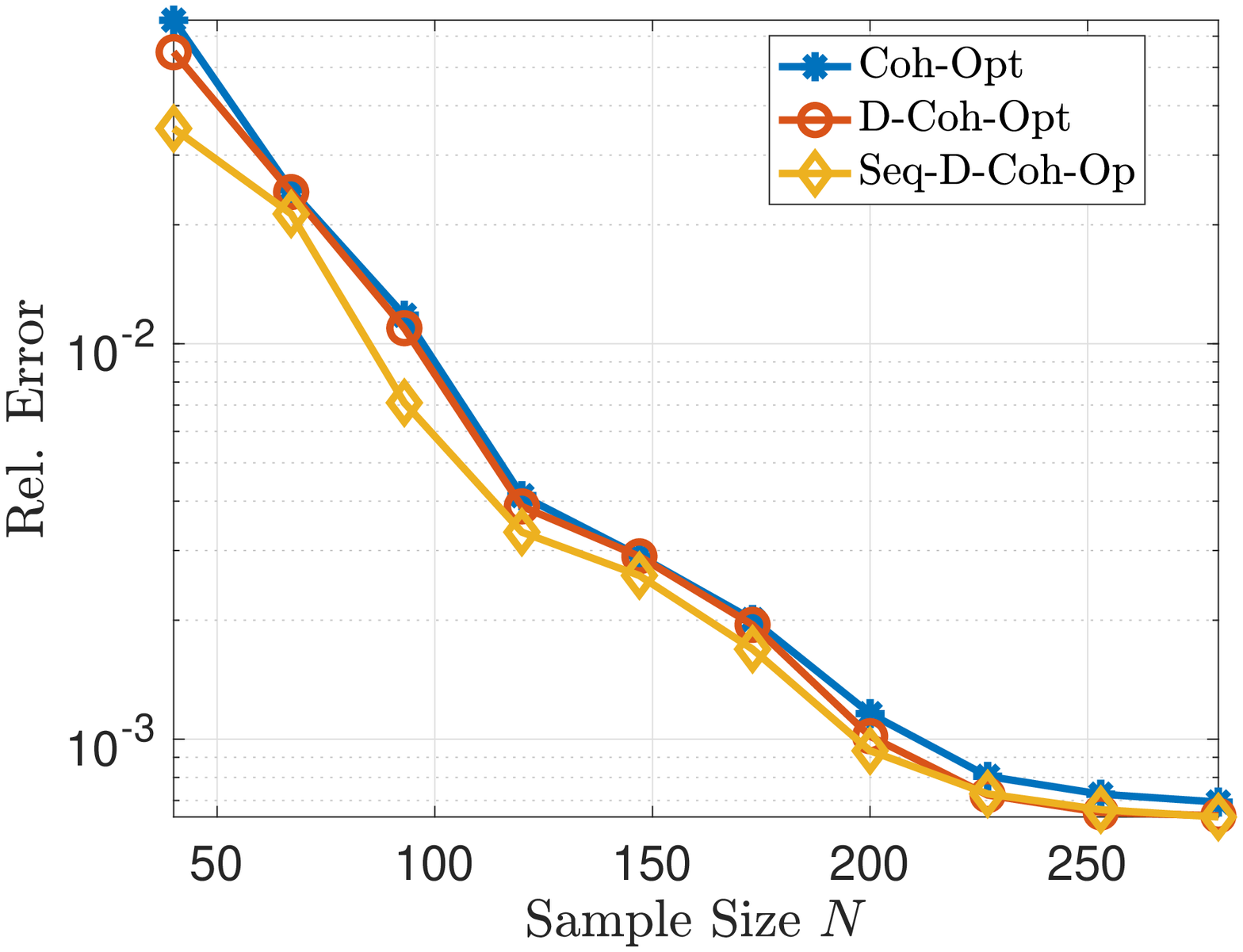}
\label{subfig:ww_rel_err_p3}
}
\subfloat[]{%
\includegraphics[width=0.49\textwidth]{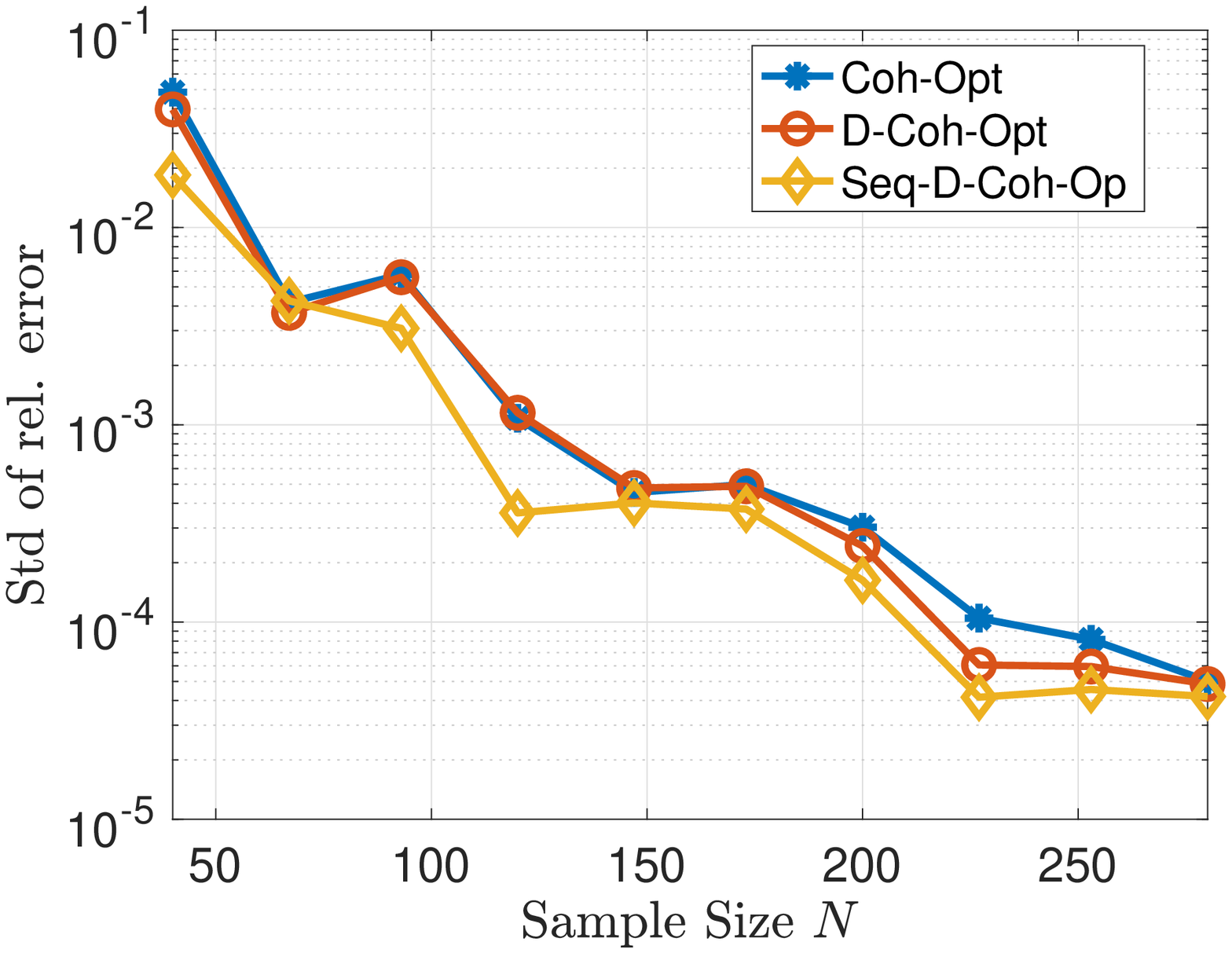}
\label{subfig:ww_std_p3}
}
}	  
\caption{The mean and standard deviation of the relative error in estimating the wing weight $u(\vXi)$ with a 3rd order PCE are shown in Figure \protect\subref{subfig:ww_rel_err_p3} and \protect\subref{subfig:ww_std_p3}, respectively}
\label{fig:ww}
\end{figure}

In this example we choose $p=3$ which corresponds to $P=286$ basis functions. For any given $N$,  let $M = 10P$ be the size of the candidate pool and $R=1000$. Figure \ref{fig:ww} depicts the mean and standard deviation of the relative validation errors. We see that Seq-D-Coh-Opt produces the most accurate PC approximations with the least variability, and that D-Coh-Opt performs only marginally better than Coh-Opt. These results seem to agree with the findings of Sections \ref{subsec:doe_testing} and \ref{subsec:msf}. 

\subsection{The Ishigami Function}\label{subsec:ishigami} 

\begin{figure}[H]    
\centering{
\subfloat[]{%
\includegraphics[width=0.49\textwidth]{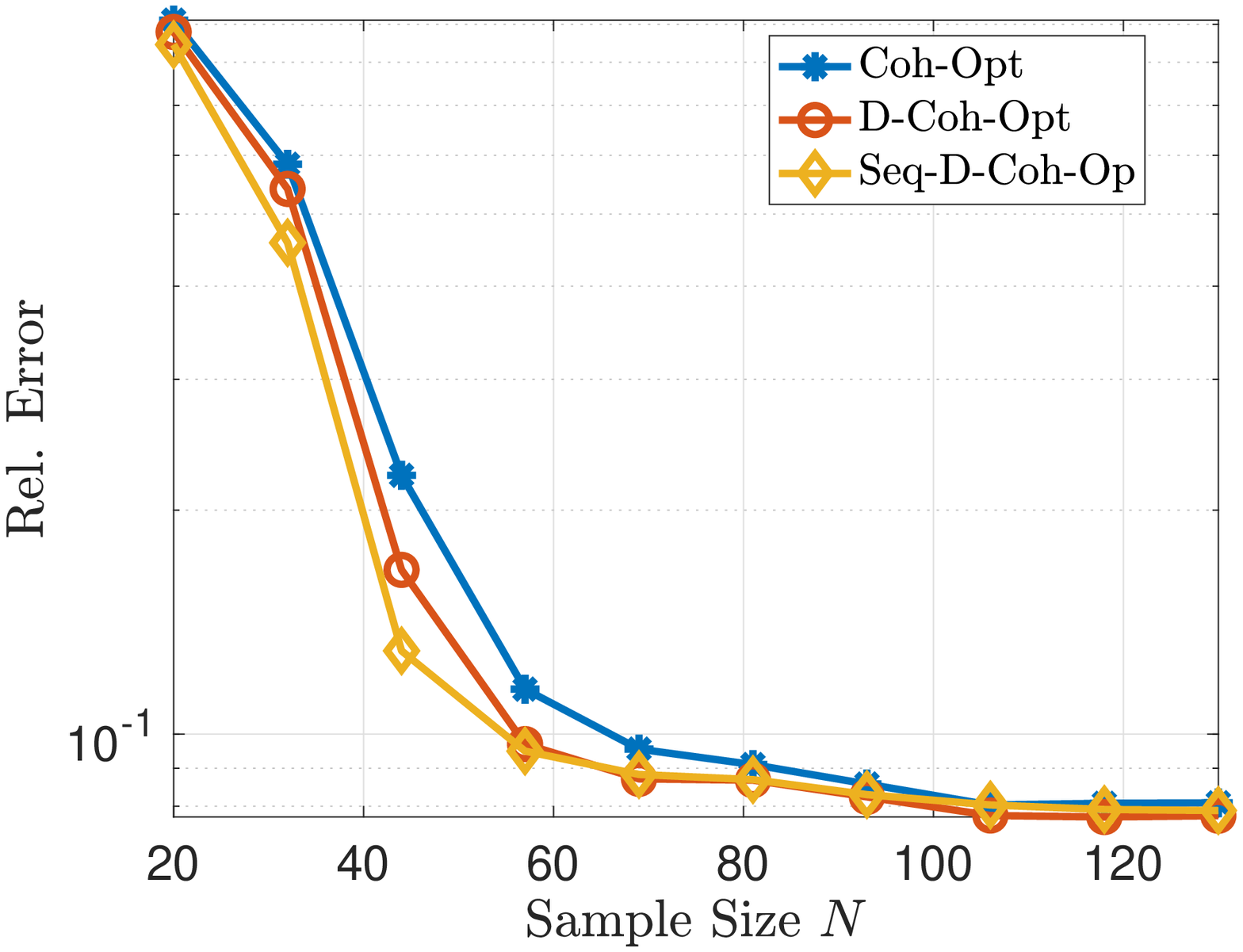}
\label{subfig:ishigami_rel_err_p7}
}
\subfloat[]{%
\includegraphics[width=0.49\textwidth]{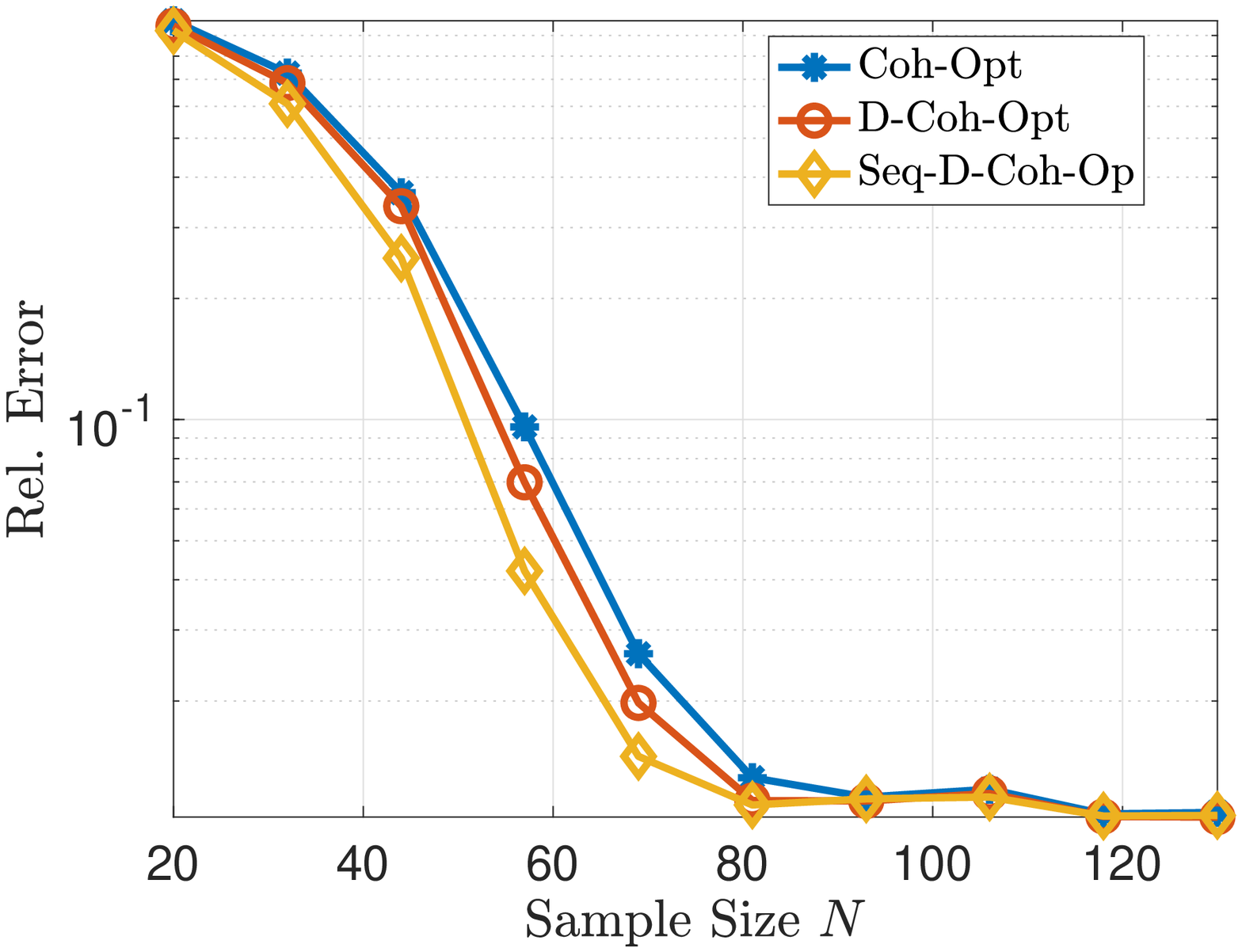}
\label{subfig:ishigami_rel_err_p9}
}\\
\subfloat[]{%
\includegraphics[width=0.49\textwidth]{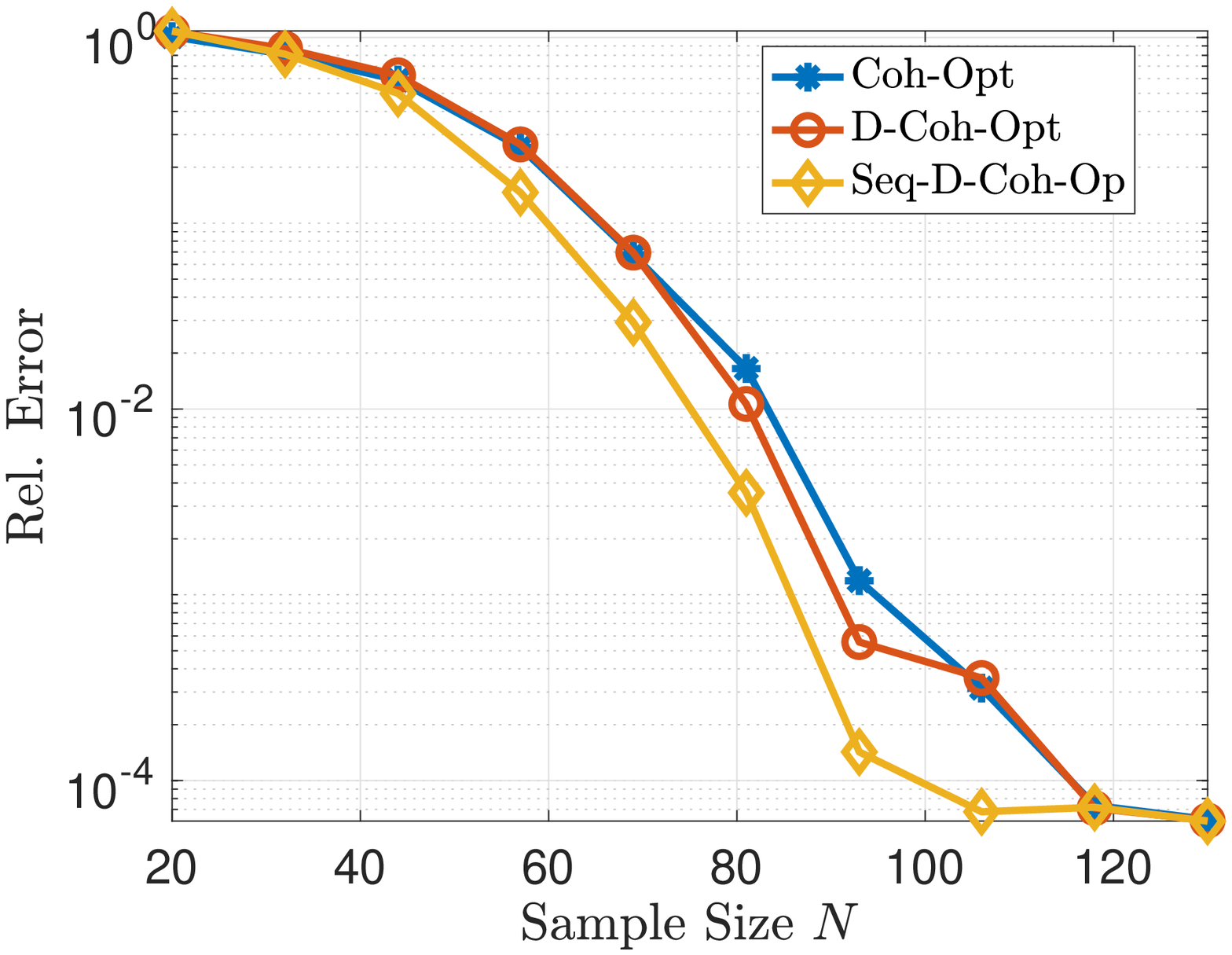}
\label{subfig:ishigami_rel_err_p12}
}
}	  
\caption{Mean of the relative error in estimating the Ishigami function with a 7th, 9th, and 12th order PC expansion in \protect\subref{subfig:ishigami_rel_err_p7}, \protect\subref{subfig:ishigami_rel_err_p9}, and \protect\subref{subfig:ishigami_rel_err_p12}, respectively.}
\label{fig:ishigami_rel_err}
\end{figure}

\begin{figure}[H]    
\centering{
\subfloat[]{%
\includegraphics[width=0.49\textwidth]{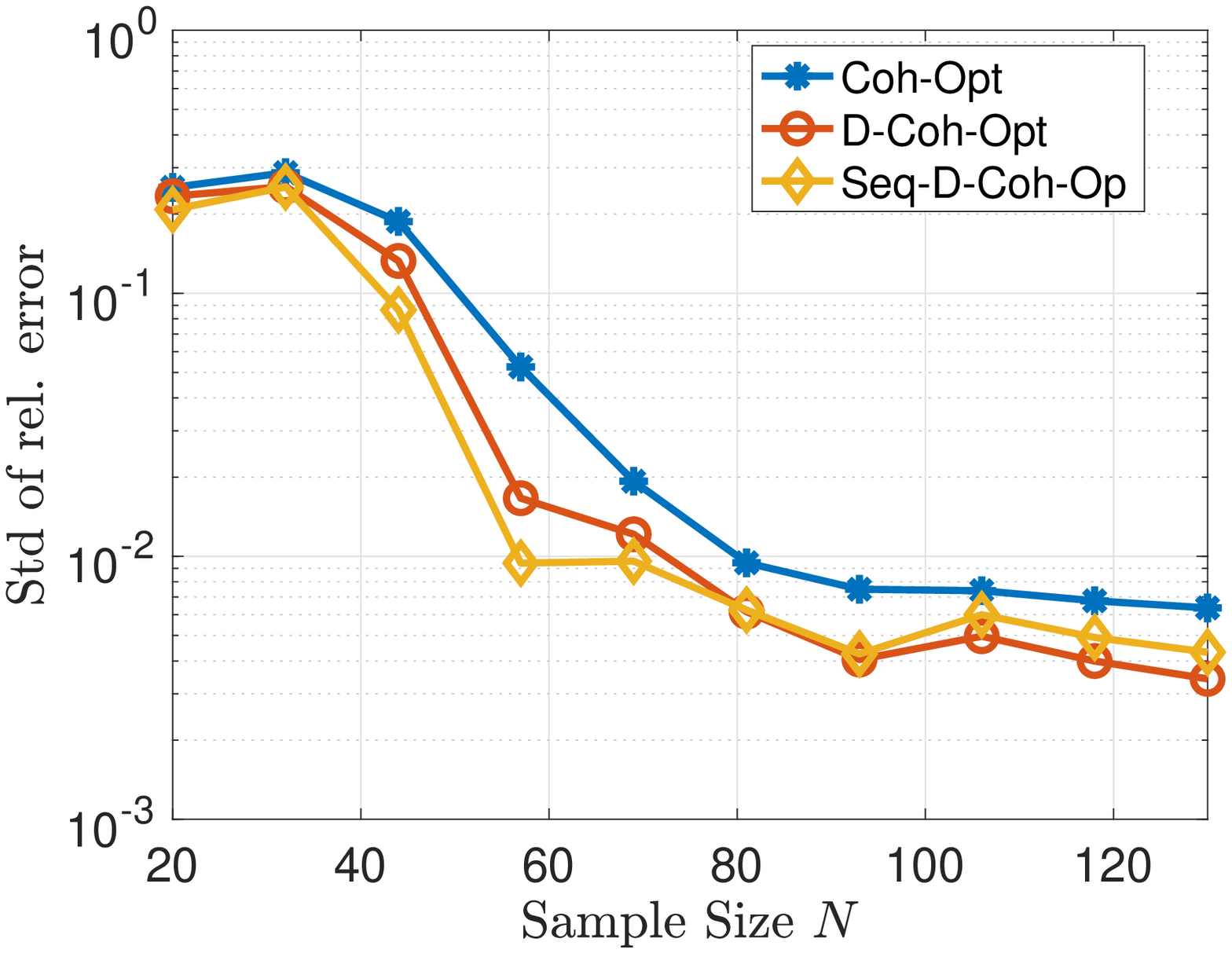}
\label{subfig:ishigami_std_p7}
}
\subfloat[]{%
\includegraphics[width=0.49\textwidth]{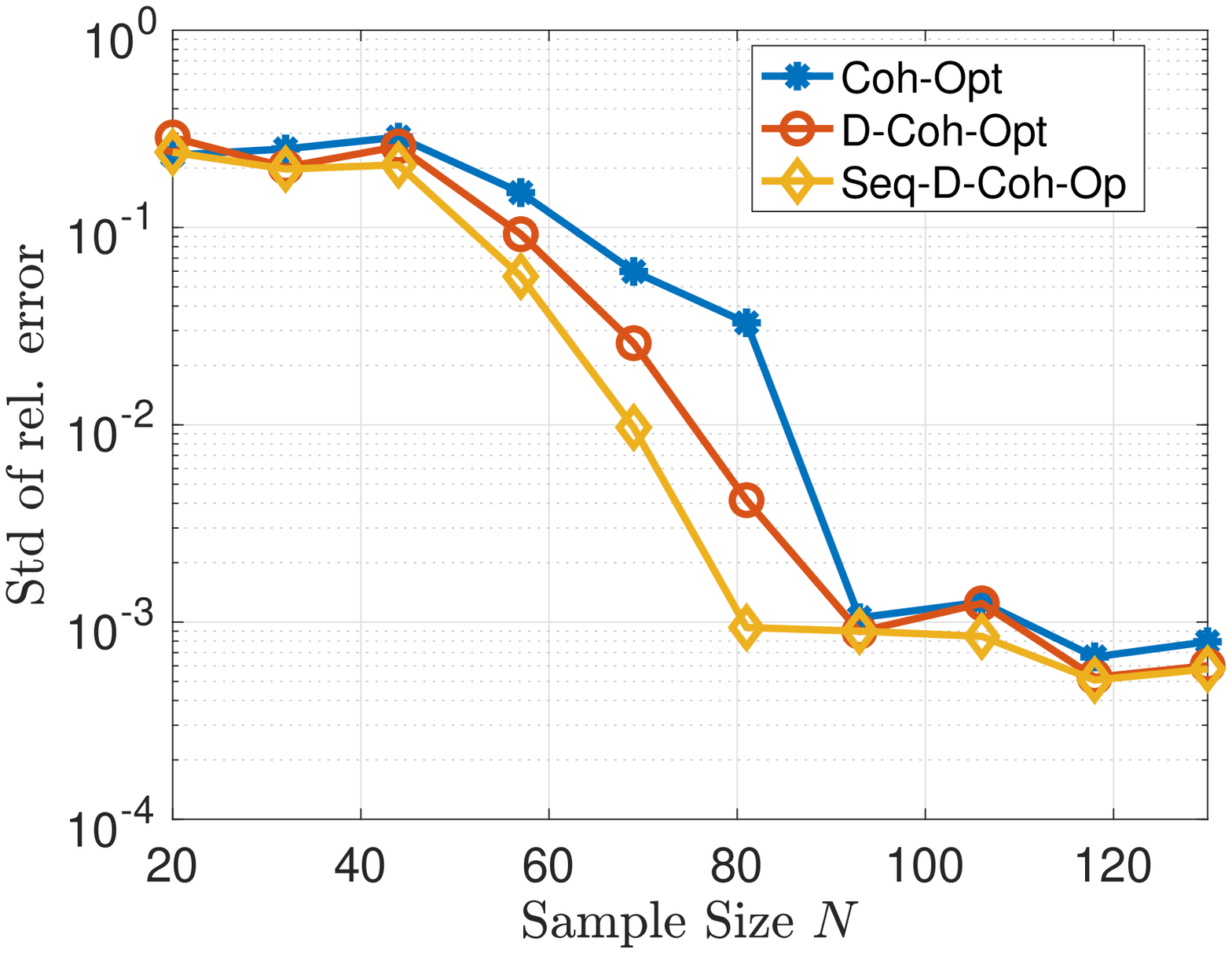}
\label{subfig:ishigami_std_p9}
}
\\
\subfloat[]{%
\includegraphics[width=0.49\textwidth]{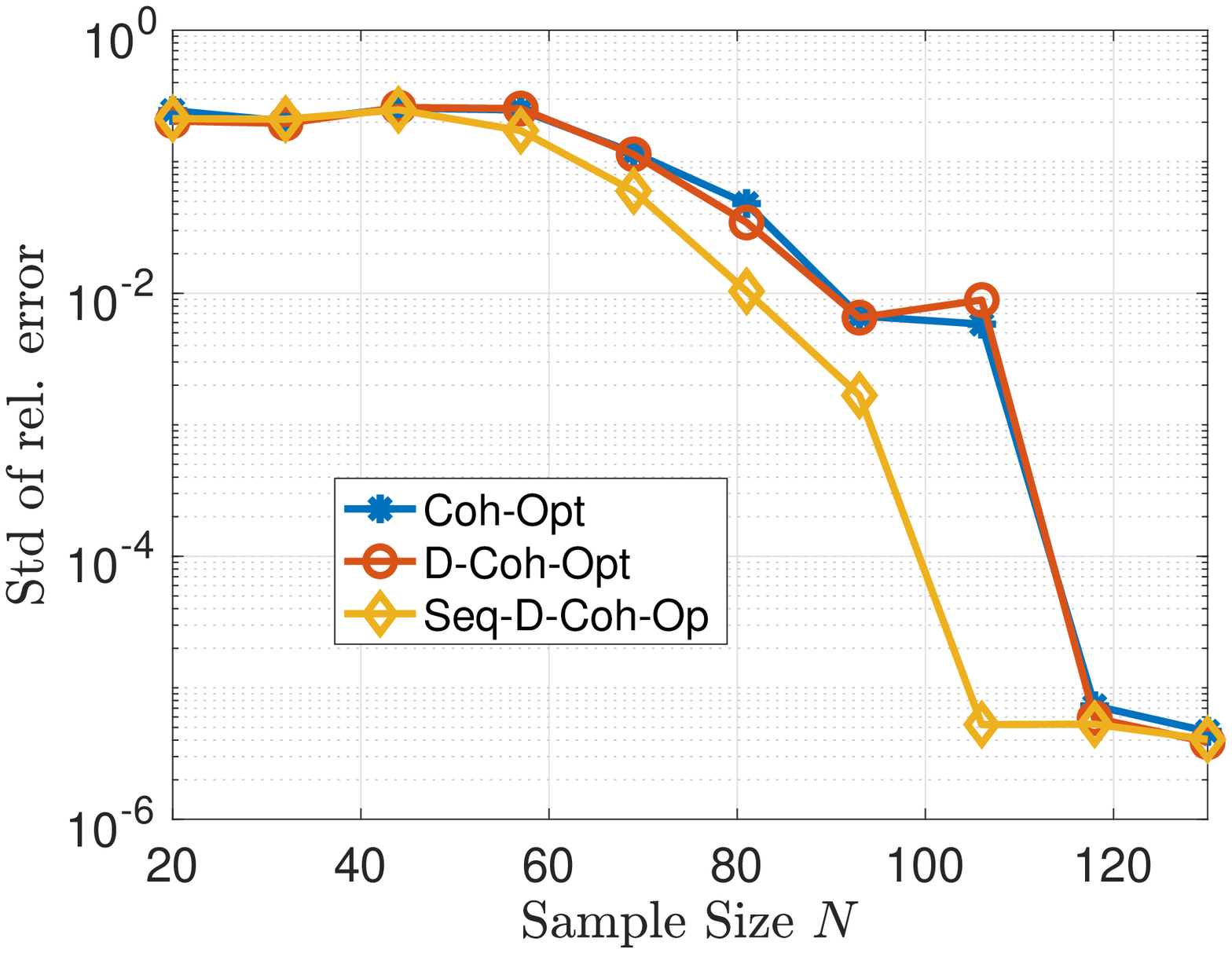}
\label{subfig:ishigami_std_p12}
}
}	  
\caption{Standard deviation of the relative error in estimating the Ishigami function with a 7th, 9th, and 12th order PC expansion in \protect\subref{subfig:ishigami_std_p7}, \protect\subref{subfig:ishigami_std_p9}, and \protect\subref{subfig:ishigami_std_p12}, respectively.}
\label{fig:ishigami_std}
\end{figure}

The last example we consider is the Ishigami function which is a well-studied benchmark problem for PC expansions \cite{ishigami1990importance,fajraoui2017optimal}. This is a 3-dimensional function with a non-monotonic and highly non-linear analytic representation given by
\begin{equation}\label{eq:Ishigami	}
f(\vXi) =  \sin x_1 + a \sin^2x_2 + bx_3^4\sin x_1, 
\end{equation}
where $x_i = \pi \Xi_i,$ and $\Xi_i \overset{\text{i.i.d.}}{\sim}  U(-1,1)$ for  $i=1,2,3$. For this example we fix the parameter values such that $a=7$ and $b=0.1$. Like the example presented in Section \ref{subsec:duffing}, this is a low-dimension, high-order example with 
$d=3$ and $p=7,9,$ and $12$ corresponding to values of $P = 120, 220,$ and $455$, respectively. For any given $N$,  let $M = 10P$ be the size of the candidate pool, and $R=1000$. Figures \ref{fig:ishigami_rel_err} and \ref{fig:ishigami_std} demonstrate the mean and standard deviation of the relative error in predicting the validation data, respectively.

We see that D-Coh-Opt and Seq-D-Coh-Opt clearly result in more accurate PC coefficient approximations, with less variability, compared to Coh-Opt, particularly for larger $p$. Further, these figures demonstrate that Seq-D-Coh-Opt results in more accurate PC coefficient approximations than the non-sequential D-Coh-Opt strategy. These results are in agreement with those presented in Sections \ref{subsec:doe_testing}-\ref{subsec:ww}, but they show a greater improvement using Seq-D-Coh-Opt versus D-Coh-Opt than in Section \ref{subsec:duffing}.  

\section{Conclusions and summary}

This work focuses on an ODE-based sampling strategy to improve the accuracy of sparse polynomial chaos expansions, given a fixed computational budget. We propose DSP, a novel sequential design, greedy algorithm for sparse PC approximation. The coefficients of the PC expansion are computed via an over-determined LSA which is a key step in the iterative SP and DSP algorithms. DSP incorporates the $D$-optimality criterion from ODE, which is known to improve the stability of PC coefficient approximation. The $D$-optimality criterion is exploited to iteratively optimize the over-determined least squares problem intrinsic to SP. This optimization leverages the estimated support set $\sS$ of the coefficient vector, which is computed at each iteration of SP. $D$-optimal designs are constructed and adapted using a RRQR algorithm. The RRQR algorithms are easier to implement than conventional algorithms for constructing $D$-optimal designs such as sequential greedy algorithms or exchange algorithms, and for an $M\times K$ matrix with $M> K$ a RRQR factorization may be computed with $\mathcal{O}(M K^2)$ floating-point operations in the worst case scenario \cite{gu1996efficient}. 

 We compare the sequential DSP algorithm to standard SP with two non-sequential strategies numerically, the first consists of purely random sampling, and the second constructs a $D$-optimal design. This comparison is done by investigating manufactured sparse PC expansions, a mathematical model for a Duffing oscillator, a wing weight function, and the Ishigami function. Additionally, a comparison between using standard Monte Carlo and coherence-optimal sampling to construct candidate design pools was conducted and the quality of these candidate pools was assessed in terms of the $D$-optimal designs they produce. Of the three physical models, the Duffing oscillator and Ishigami function represent low-dimension, high-order cases, while the wing weight function represents a high-dimension, low-order case. For the manufactured sparse PC expansions both low-dimension, high-order and high-dimension, low-order cases are considered. Further, for manufactured sparse PC expansions of low-dimension, high-order a comparison of standard Monte Carlo sampling to coherence-optimal sampling is performed. Our results support the following conclusions.
 \begin{itemize}
 \item  Coherence-optimal sampling results in better experimental designs, and more stable and accurate PC coefficient approximations as compared to standard Monte Carlo sampling.
 \item  The SP algorithm using sequential $D$-coherence-optimal sampling (Seq-D-Coh-Opt) outperforms the non-sequential strategies (D-Coh-Opt and Coh-Op), in terms of providing more accurate PC approximations with less variability for a fixed computational budget. 
 \item The greatest improvements of Seq-D-Coh-Opt (and D-Coh-Opt) are for low-dimension, high-order PC expansions.
 \item RRQR factorizations can be used to efficiently construct and update $D$-optimal designs.
 \end{itemize}
 Our future work includes investigating other greedy algorithms for sparse signal reconstruction, such as OMP or CoSaMP, and other alphabetic optimality criteria for design adaptation.

\begin{appendices}
\section{DSP Algorithm with Cross-validation for K}

The following modified DSP algorithm is employed to approximate the PC coefficients for the models discussed in Section \ref{sec:numerics}. This algorithm is similar to Algorithm \ref{alg:DSP}, except it uses Algorithm \ref{alg:cross_val_K} at each iteration to estimate the optimal value of $K$. 
\begin{algorithm}                      
\caption{$D$-optimal Subspace Pursuit (DSP) with cross-validation on $K$}          
\label{alg:DSP_cross_val}                           
\small
\textbf{Input:} $\mPhi_c, N_{max}$\\
\textbf{Initialization:} \\
 1) Let $N_0 =\lfloor 0.8 N_{max} \rfloor$ \\
 2) $\mPhi_{N_0} = \mPhi_c( \vpi_{N_0}, \; : \; )$ corresp. to $\braces{\vxi_i}_{i=1}^{N_0}$ , and $\vv^0 = [ w(\vxi_1)u(\vxi_1),\ldots,w(\vxi_{N_0})u(\vxi_{N_0})]^T$. \\
3) Estimate $K$ according to Algorithm \ref{alg:cross_val_K} with $\mPhi_{N_0}$ and $\vv^0$. \\
4) $\sS^0 = \left\lbrace K \text{ indices corresp. to the largest in } |\cdot | \text{ entries of the vector } \mPhi_{N_0}^T\vv^0   \right\rbrace$. \\ 
5) $\vv_r^0 = \text{resid}(\vv^0,\mPhi_{N_0}(\;:\;,\sS^0)).$\\  

\textbf{Iteration:} At the $\ell$th iteration,  let $N$ = length$(\vv^{\ell-1})$ and perform the following steps:\\
1) If $\ell>1$ and $N \neq$ length$(\vv^{\ell -2})$, then set $K$ according to Algorithm \ref{alg:cross_val_K} with $\mPhi_{N}$ and $\vv^{\ell-1}$. \\
2) $\tilde{\sS}^{\ell} = \sS^{\ell -1}\cup \lbrace K$ indices corresp. to the largest in $|\cdot |$ entries of the vector $\mPhi_{N}^T \vv_r^{\ell-1} \rbrace$.\\
3) Set $\hat{\vc} = \mPhi_{N_\ell}(\;:\;,\tilde{\sS}^{\ell}) ^\dagger\vv^{\ell-1}$.\\
4) $S^\ell = \lbrace K$ indices corresponding to the largest in $|\cdot |$ elements of $\hat{\vc} \rbrace$.\\
5) If $N < N_{max}$, then $\mPhi_{N+1} = \mPhi_c(\vpi_{N+1},\; : \;)$ and $\vv^\ell = [\vv^{\ell-1} \; ; \; w(\vxi_{N_\ell+1})u( \vxi_{N_\ell+1} )]$. \\
6) $\vv_r^\ell = \text{resid}(\vv^\ell,\mPhi_{N+1}(\;:\;, \sS^\ell))$. \\
7) If $N = N_{max}$ and $\ell = P$, quit iterating.\\
8) If $ ||\vv_r^\ell || > || \vv_r^{\ell -1}||$  and $\ell \geq N_{max} -N_0 +1$, set $\sS^\ell = \sS^{\ell-1}$ and quit iterating. \\ 

\textbf{Output:} \\
1) The approximate PC coefficients $\hat{\vc}$,  satisfying $\hat{\vc}(\braces{1,\ldots,P}\setminus \sS^\ell) = \bm{0}$ and $\hat{\vc}(\sS^\ell) = \mPhi_{N_{max}}(\;:\;,\sS^\ell)^\dagger \vv^\ell$.
\end{algorithm}
\section{Proof of Theorem \ref{thm:stochastic_dominance}}\label{appendix:proofs}

\renewcommand{\theequation}{\thesection.\arabic{equation}}
\setcounter{equation}{0}

\begin{proof}
Recall that, in this case, coherence-optimal sampling is the sampling that minimizes the coherence $\mu_K$ over all independent importance sampling distributions \cite{hampton2015coherence,hampton2015compressive}. First, we restate a bound given by \cite[Lemma~5.1]{hampton2015coherence}, in some increased generality, and ignoring contributions from domain truncation for a more direct exposition. This result relates to bounds in \cite{cohen2013stability,tropp2012user,vershynin2010introduction}, and for $t\in (0,1)$ gives a Chernoff-type bound in the form of
\begin{equation}\label{eq:chernoff_type_bound}
\Prob{\|\mM-\mI\| \geq t }\leq 2K\exp\parens{-cM\mu_K^{-1}t},
\end{equation}
where in this case, $\mM \in \mathbb{R}^{K\times K}$ is constructed from a design matrix utilizing all $M$ samples in a candidate pool and $K$ is the total number of basis functions used. If $t<1$ and $\sigma_k$ denotes the singular values of $\mPhi_c$, then $\| \mM - \mI\|\leq t$ implies that for all $k$,
\begin{equation}
\vert \sigma_k(\mM) -1 \vert \leq t,
\end{equation}
which in turn implies that 
\begin{equation}
\sigma_k(\mM) \geq 1-t.
\end{equation}
Note that $K$ denotes the rank of $\mM$, and that
\begin{equation}
\vert \det\mM \vert = \prod_{k=1}^K\sigma_k(\mM)\geq (1-t)^K.
\end{equation}
Finally, we mention that with some algebraic manipulation we can present the Chernoff bound in \eqref{eq:chernoff_type_bound} in terms of $\phi_D$ in \eqref{eq:phi_D_bound}. 
\end{proof}
\end{appendices}
\newpage
\bibliographystyle{elsarticle-num}
\bibliography{dopt_pce}
\end{document}